\documentclass{aa}

\usepackage{graphicx}
\usepackage[normalem]{ulem}
\usepackage{txfonts}

\usepackage{hyperref}  
\hypersetup{
    colorlinks=true,
    linkcolor=blue,
    filecolor=magenta,
    citecolor=blue,
    urlcolor=blue,
}
\usepackage{natbib}
\bibpunct{(}{)}{;}{a}{}{,}
\usepackage{multirow}
\usepackage{nicefrac}
\usepackage[usenames,dvipsnames]{color}
\usepackage{caption}
\usepackage{subcaption}
\usepackage[normalem]{ulem}

\newcommand{\kms}{km\,s$^{-1}$}
\newcommand{\ms}{m\,s$^{-1}$}

\def\specchar#1{{\sc #1}}
\def\FeI{\mbox{Fe\,\specchar{i}}}
\def\FeII{\mbox{Fe\,\specchar{ii}}}
\def\SrII{\mbox{Sr\,\specchar{ii}}}
\def\CaII{\mbox{Ca\,\specchar{ii} IR}}

\def\HeI{\mbox{He\,\specchar{i}}}
\def\HeD3{\mbox{He\,\specchar{i}\,\,D$_3$}}
\def\NaD{\mbox{Na\,\specchar{d}}}
\def\NaD2{\mbox{Na\,\specchar{d}$_2$}}
\def\SrII{\mbox{Sr\,\specchar{ii}}}
\def\CaIIK{\mbox{Ca\,\specchar{ii}\,\,K}}       
\def\CaIIH{\mbox{Ca\,\specchar{ii}\,\,H}}       
\def\Halpha{\mbox{H$\alpha$}}
\def\Hbeta{\mbox{H$\beta$}}
\def\Hepsilon{\mbox{H$\epsilon$}}
\def\Hgamma{\mbox{H$\gamma$}}
\def\Hdelta{\mbox{H$\delta$}}

\definecolor{lena}{rgb}{0.0,0.5,0.0}

\begin{document}

\title{Two fluid dynamics in solar prominences}
\author{S. J. Gonz\'alez Manrique\inst{1,2,3,4}, E. Khomenko\inst{2,3},
M. Collados\inst{2,3}, C. Kuckein\inst{2,3}, T. Felipe\inst{2,3}, P. G{\"o}m{\"o}ry\inst{4}
}

\authorrunning{Gonz\'alez Manrique et al.}

\institute{%
    $^1$ Leibniz-Institut für Sonnenphysik (KIS), Schöneckstr. 6, 
         79104 Freiburg, Germany \\
    $^2$ Instituto de Astrof\'{i}sica de Canarias (IAC), 
         V\'{i}a L\'{a}ctea s/n, 38205 La Laguna, Tenerife, Spain\\  
    $^3$ Departamento de Astrof\'{\i}sica, Universidad de La Laguna
         38205, La Laguna, Tenerife, Spain \\ 
    $^4$ Astronomical Institute, Slovak Academy of Sciences (AISAS), 
         05960 Tatransk\'{a} Lomnica, Slovak Republic\\
    \email{smanrique@leibniz-kis.de}}

\date{Received September 30, 2023; accepted November 5, 2023}

\abstract{
}
{
Solar prominences contain a significant amount of neutral species. The dynamics of the ionised and neutral fluids composing the prominence plasma can be slightly different if the collisional coupling is not strong enough. This differential dynamics can be discerned by tracing line of sight velocities using observational techniques. Large-scale velocities can be quantified by measuring the global local and instantaneous displacement of spectral lines by Doppler effect. Small-scale velocities leave their imprint on the width of spectral lines. In addition, these small-scale velocities can have a thermal (pure stochastic motion) nature or a non-thermal (small-scale unresolved instabilities, high-frequency waves, etc.) origin.  Here we use one spectral line of ionised and two spectral lines of neutral elements to measure the resolved and unresolved velocities in a prominence with the aim to investigate the possible decoupling of the observed charged and neutral species.}
{
A faint prominence was observed with the German Vacuum Tower Telescope (VTT) on June 17, 2017. Time series consisting of repeated 10-position scans over the prominence were performed while recording simultaneously the intensity spectra of the \CaII\ 854.2 nm, \Halpha\ 656.28 nm, and \HeD3\ 587.56 nm lines. The line of sight velocities and the Doppler width of the three spectral lines were determined at every spatial position and temporal moment. To make sure all spectral lines were sampling the same plasma volume, we applied selection criteria to identify locations with optically thin plasma. In addition, asymmetric or double-peaked profiles were also excluded for the analysis, since (even in an optically thin regime) they are indicative of the presence of strong velocity gradients or multiple components in the line of sight. Thus, only optically-thin, symmetric, single-lobed profiles were retained for this study. As an additional reliability test of the selection criteria, we have also compared our results with optical thickness calculations. }
{
After the application of all the selection criteria, only a region close the prominence border met all requirements. The velocities of the three spectral lines turned out to be very similar over this region, with the ionised \CaII\ showing velocity excursions systematically larger compared to those of the neutral lines of \Halpha\ and \HeI\ at some moments. The latter were found to be much closer to each other. Most of the velocity differences were below 1 \kms. The analysis of the Doppler widths indicated that the \CaII\ line shows an excess of unresolved motions. We cannot establish whether these velocities are related to a different temperature of the ions or to unresolved small-scale motions due to any non-thermal mechanism.  
}
{
The dynamics of the ionised and neutral plasma components in the observed prominence was very close one to the other. The differences found may indicate that a localised decoupling between ions and neutrals may appear at particular spatial locations or instants of time. Indications of different unresolved motions between those species have also been obtained.
}


\maketitle

\section{Introduction}\label{Sect:Introduction}

The solar atmosphere is composed of a mixture of ionised and neutral gases. In most situations, the charged and neutral components follow the same dynamics, as a consequence of the very strong collisional coupling. Nevertheless, it is not always the case. Many theoretical studies have shown that, as the collisional coupling weakens with height in the solar atmosphere (mainly due to the fast decrease of the density), neutrals and charges can move with different velocities, leading to a dynamical decoupling and frictional heating between the components \citep{Khomenko2017,Ballester2018}. While the number of theoretical studies of the effects of charge-neutral interaction (generically called ``partial ionisation effects'') in the solar atmosphere has been increasing, there are not so many observational confirmations of the presence of these effects in the Sun. Direct observations of the decoupling between the plasma components, and the associated effects, are extremely challenging, since, as theory has shown, they are expected to occur on rather small spatial and temporal scales. Fortunately, the relevant scales for partially ionised plasmas are still much larger than for the fully ionised plasmas, since they are related to ion-neutral, and not ion-electron collisional scales. Theoretical models and simulations suggest values below $\sim10^1$ km for the spatial and \hbox{$\sim10^0$ s} for the temporal scale \citep{Khomenko2012, MartinezSykora2012, Khomenko2014}.

Partial ionisation effects are expected to be more prominent in rarefied plasmas, such as the one present in the upper chromosphere of the Sun, or in transitions layers between the prominence and coronal plasmas \citep{Khomenko2014, Khomenko2021, Popescu2021a, Popescu2021b, MartinezGomez2022, MartinezSykora2023}. They are also expected to increase at locations with stronger magnetic fields, since, due to the action of the Lorentz force on the charged component, ions can experience a large acceleration and, consequently, reach larger velocities compared to neutrals \citep{Khomenko2014, Popescu2019}. For example, simulations of fast magneto-acoustic waves propagation in the solar chromosphere in the two-fluid approach by \citet{Popescu2019} revealed slightly larger amplitudes for the ionised component. A similar behaviour was also observed by \citet{Zhang2021} in their two-fluid simulations of slow acoustic shocks. The difference between the ion and neutral velocities is especially large at discontinuities such as shock fronts, and increases with the wave amplitude. Shock fronts reveal a complex multi-fluid structure in partially ionised plasmas \citep{Draine1993, Hillier+Snow2023}. Even for sub-sonic flows, a large decoupling of the order of 1 \kms\ between the ionised and neutral components has been found at the prominence-corona transition layer in simulations of the Rayleigh-Taylor instability (RTI) in a solar prominence thread by \citet{Popescu2021a, Popescu2021b}, and in regions surrounding coronal rain drops, as revealed by the simulations of \citet{MartinezGomez2022}.

Observationally, there are a number of works where the existence of ion-neutral decoupling might be behind the observed behaviour, or, at least, it can be one of the possible explanations. For example, \citet{Gilbert2007} found observational indications for the cross-field diffusion of neutral material in filaments and prominences, a mechanism earlier suggested by \citet{Gilbert2002}. These authors found a temporal change of the relative Hydrogen to Helium abundance in the upper parts of prominences compared to their lower parts. One of the possible explanations is a slightly faster drainage of Helium compared to Hydrogen caused by the difference in their atomic mass. Another indirect evidence of the presence of partial ionisation effects comes from measurements of a possible misalignment between the magnetic field and some chromospheric \Halpha\ fibrils \citep{delaCruzRodriguez2011, AsensioRamos2017}. Numerical simulations of solar magneto-convection including ambipolar diffusion as the main partial ionisation effect in the single fluid approximation \citep{MartinezSykora2016} have shown that a cross-field plasma motion caused by the neutral drag can produce a visible misalignment between the magnetic field inclination and the structure observed in density (a fibril). A strict alignment is expected from ideal magneto hydrodynamics, since plasma flows are forced to follow the magnetic field lines. In the same vein, slight differences in the speeds of the Evershed flow in sunspots detected from neutral \FeI\ and ionised \FeII\ lines were reported by \citet{Khomenko2015}. At all radial distances in the penumbra, the \FeII\ velocities were a few hundred m/s larger than the corresponding \FeI\ velocities, with the difference increasing with height. A possible explanation is that the strong magnetic field of sunspots might lead to stronger decoupling effects, detectable even in the photosphere. 

Recently, many efforts have been dedicated to directly measure spatially and temporally resolved velocities of ions and neutrals. For that, several works have focused on solar prominences \citep{Khomenko2016, Anan2017, Wiehr2019, Wiehr2021, Zapior2022}. These targets have several advantages: (1) prominences are observed at the limb, where the chances to have an optically-thin plasma are larger, to guarantee that the spectral lines of different elements are formed over the same plasma volume; (2) numerical simulations suggest that a strong ion-neutral decoupling can take place at the transition region between prominences and corona \citep{Popescu2021a, Popescu2021b, MartinezGomez2022}; (3) prominences show a wide variety of dynamical effects, such as waves, instabilities or flows, whose properties can be modified by partial ionisation effects. 

\citet{Khomenko2016} measured velocities using the ionised \CaII\ 854.2 nm and the neutral \HeI\ 1083 nm using a very high cadence time series of spectra of a prominence. These authors performed a Principal Component Analysis (PCA) classification of the profiles to exclude those with irregular shapes, as they were indicative of the presence of multiple components in the line of sight. In addition, they used the relative amplitudes of the blue and red components of the \HeI\ 1083 nm triplet to detect and exclude locations with large optical thickness. In the remaining locations, the velocities from the ionised and neutral lines turned out to be very similar. Still, at locations with large flows and large spatial gradients these authors found slightly larger velocities of the ionised \CaII\ compared to that of the neutral \HeI. Unlike that, \citet{Anan2017} obtained fixed slit high-cadence prominence spectra of a larger set of spectral lines (\hbox{\Hepsilon\ 397 nm}, \hbox{\Hgamma\ 434 nm}, \hbox{\CaIIH\ 396.8 nm}, and \hbox{\CaII\ 854.2 nm}) and concluded that differences between the velocities derived from any pair of these spectral lines were of the same order of magnitude, regardless whether ion-ion, ion-neutral or neutral-neutral species were used for the comparison. These authors used \Hepsilon\ amplitudes to determine locations with optically thin plasma. \citet{Wiehr2019} observed in a quiescent prominence using spectral lines of ionised \SrII\ 407.8 nm and neutral \NaD2\ 589.0 nm, together with the nearby lines of \HeI\ 501.5 nm and \FeII\ 501.8 nm, and, separately, \Hdelta\ at 410.1 nm. All the spectral lines were in the optically thin regime, as verified by \Hdelta. These authors found a systematic excess of the velocities of ions by about 10-20\% at the locations with the largest velocities, both for the \SrII/\NaD2\ and \FeII/\HeI\ line pairs. In a subsequent work, \citet{Wiehr2021} detected an even larger, 20-25\%, ion velocity excess in the \FeII/\HeI\ line pair, which translates into an up to 300-700 \ms\ absolute velocity difference. The locations with an ion velocity excess were seen to be restricted in space and time to about 5 Mm and 5 min, respectively. An even larger ion velocity excess, up to 70\%, was detected at locations with high-frequency oscillations of 22 sec period. Finally, \citet{Zapior2022} reported observations of a scan of a large dense prominence using six simultaneous spectral lines: \CaIIH, \Halpha, \Hbeta, \Hepsilon, \HeD3\ and the \CaII. After the analysis of the optical thickness at the various locations of the prominence, only a small optically thin area at its border was retained where, again, larger ionised Calcium velocities were obtained compared to those of neutral Hydrogen. Their observations also showed much smaller differences between the velocities calculated with spectral lines of the same element. 

There have also been attempts to detect differences in the kinetic temperature and unresolved motions in the ionised and neutral species that form prominences \citep{Ramelli_etal_2012, Park_etal_2013, Wiehr_etal_2013, Stellmacher_Wiehr_2015, Stellmacher_Wiehr_2017, Okada_etal_2020}. All these works lead to a range of prominence temperature between 4000 and 20000 K and non-thermal velocities from a few to up to 20 \kms. \citet{Okada_etal_2020} make an exhaustive analysis of the behaviour of typical spectral lines used in prominences and conclude that optically thick and thin lines tend to give different results and 
a special care has to be taken with the opacity. Despite all these difficulties, \citet{Ramelli_etal_2012} and \citet{Stellmacher_Wiehr_2015, Stellmacher_Wiehr_2017} find that ions tend to have a larger Doppler width than the neutral species. 

Despite the accumulated evidence, it is still hard to unambiguously attribute the observed differences in the ion and neutral (resolved and non-resolved) velocities to the existence of partial ionisation effects (i.e., to ion-neutral decoupling). In general, the observed differences in bulk velocities are of the order of hundreds of \ms, which fit well into the theoretical expectations. Theory seems to suggest that ions should move slightly faster as a consequence of the Lorentz force induced by the magnetic field and most of the observations seem to point in this direction. However, according to the numerical simulations, the velocity decoupling might only exist over very narrow areas 
\citep{Popescu2021a, Popescu2021b, MartinezGomez2022}. None of the observations reported above has enough spatial resolution to claim an unambiguous detection. In addition, all those observations lack magnetic field information, which would greatly help to attribute the measurements to real multi-fluid effects. Still, one needs to keep accumulating the statistics of measurements using multi-line observations with the highest cadence and largest spatial resolution possible. While the spatial resolution of observations is never expected to achieve the scales of the few km needed to resolve multi-fluid effects, the velocity decoupling or the different Doppler width of spectral lines might still show up in volume-averaged data, as suggested by the works mentioned above. 

In the current paper we complement the work of \citet{Khomenko2016} by measuring the resolved and non-resolved velocities of neutral and ionised species in a prominence plasma using three spectral lines (\Halpha, \HeD3 and \CaII), with the best temporal cadence allowed by the observational set-up. The paper is organised as follows: Sect. \ref{Sect:Observations} 
describes the observations and data reduction; Sect. \ref{Sect:data analysis} is devoted to the analysis of the spectral profiles, their classification according to their shape, the evaluation of the optical thickness and the methods to calculate the velocities; Sect. \ref{Sect:results} shows the results obtained, which are discussed in Sect. \ref{Sect:discus_conclus}.  

\begin{figure}[!t]
 \centering
 \includegraphics[width=\hsize]{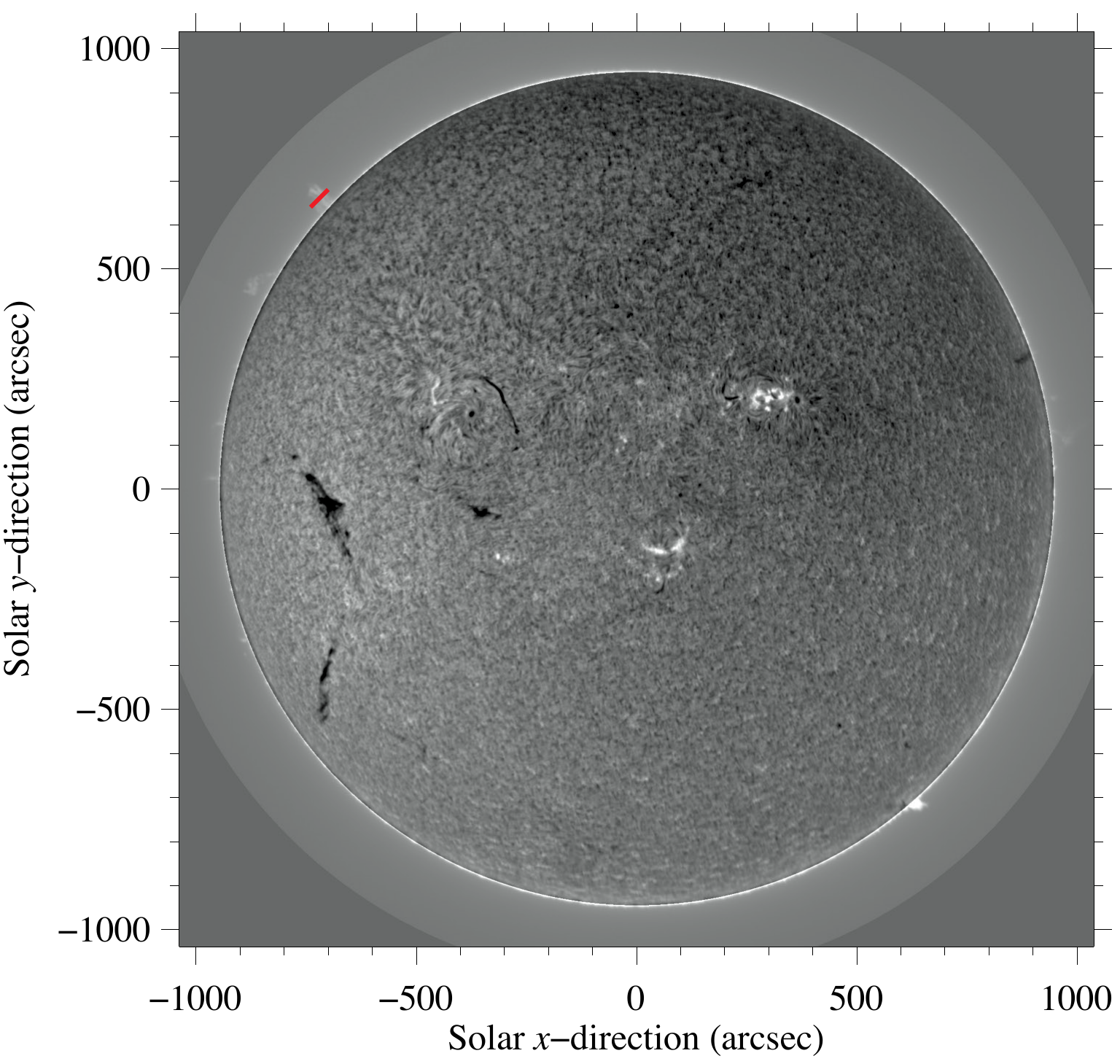} 
\caption{ChroTel \Halpha\ filtergram taken on 2017 June 17 at 09:51~UT during the spectroscopic 
observations. The red line on the north-east (upper left) limb depicts the slit position of the 
echelle spectrograph. A limb darkening correction has been applied to the image.} 
 \label{Fig:Chrotel}

 \centering
 \includegraphics[trim=60 0 60 6, width=\hsize]{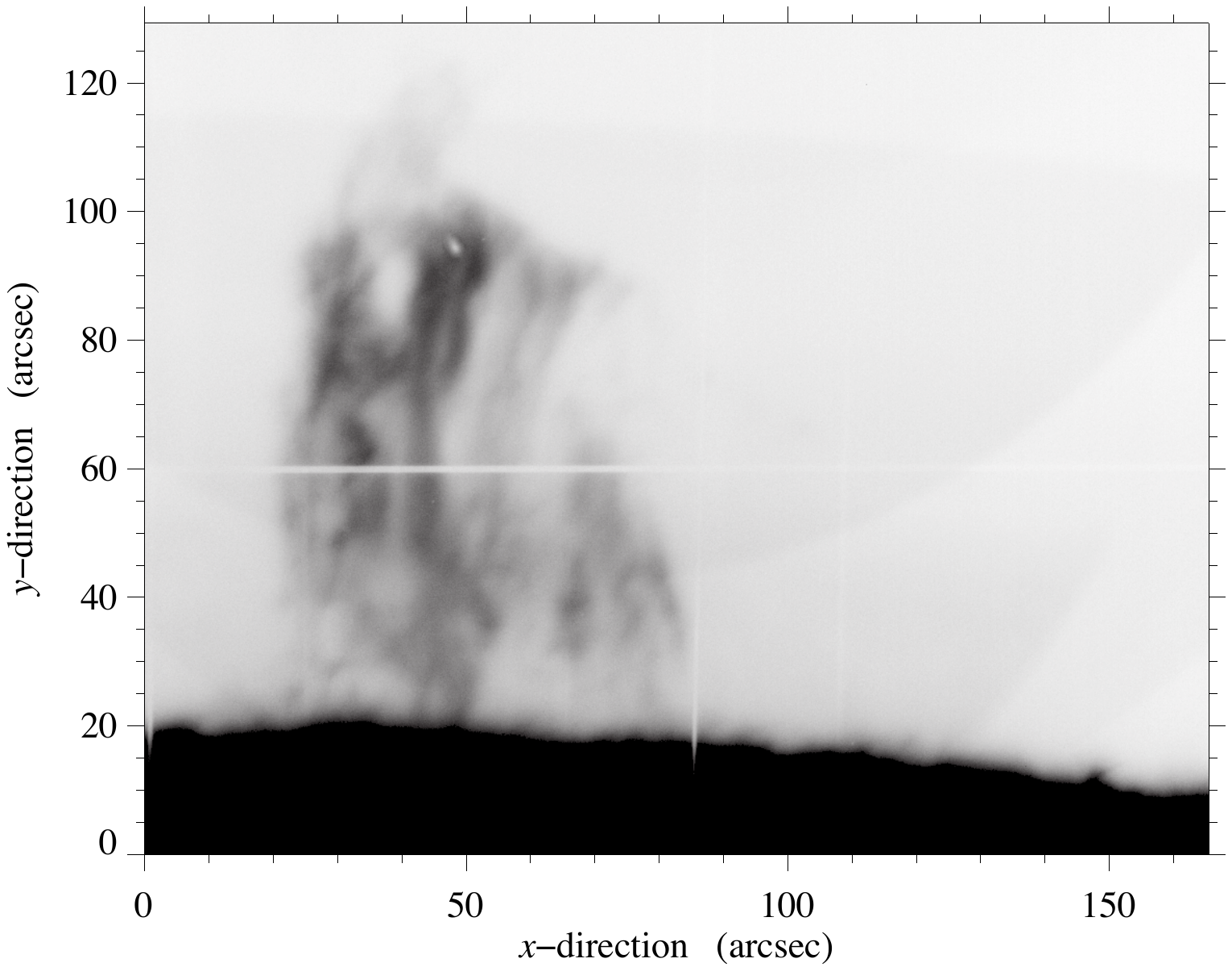} 
\caption{Slit-jaw \Halpha\ line core image. The horizontal 
white line depicts the slit position of the spectrograph 
where the observation of the three spectral lines were performed (ten adjacent positions
separated vertically 0$\farcs$75 were acquired). This image was taken 9 minutes after the 
start of the observing sequence. 
} 
 \label{Fig:SJ}
 \end{figure}

\section{Observations and data reduction}\label{Sect:Observations}
A solar prominence was observed at the eastern limb between 09:58~UT and 
11:30~UT on June~17, 2017. The prominence was located at the North-East 
off-limb at coordinates $X=-712\arcsec$ and $Y=654\arcsec$. The data used 
in this study were acquired at the German Vacuum Tower 
Telescope \citep[VTT, ][]{vonderLuehe1998}
located at the Observatorio del Teide, Tenerife. Figure \ref{Fig:Chrotel} 
displays an \Halpha\ full-disk solar image taken close to the beginning 
of our observations and acquired with the ChroTel small telescope installed 
in the same building as the main telescope \citep{Chrotel}. The observations 
were performed under relatively good seeing conditions and the Kiepenheuer 
Adaptive Optics System of the VTT \citep[KAOS, ][]{vonderLuehe2003,Berkefeld2010}
ensured a good data quality during the whole data series. The AO system was locked 
on existing small-scale photospheric structures near the limb in a stable way 
during the whole series. The recorded prominence was rather
stable, showing a slow evolution during the observations,
with the left barb evolving faster than the right one.

The \CaII\ 854.2 nm, \Halpha\ 656.28 nm, and \HeD3\ 587.56 nm spectral 
lines were simultaneously observed using the 
echelle spectrograph attached to the VTT with a slit 0$\farcs$75 wide. 
The spectrograph was used in spectroscopic mode to ensure a highest possible 
cadence and reach the signal to noise needed for this study. The same type of
camera and detector was used to record the spectral 
lines \Halpha\ and \HeD3\ (PCO-4000). Differently, a
PCO-2004 Sensicam camera was used to register the \CaII\ spectral line. 
The spatial sampling along the slit was 0$\farcs$34
for the \Halpha\ and \HeD3\ spectral lines, and
0$\farcs$244 for the \CaII\ spectral line. 
The exposure time for the three cameras was synchronised by an external trigger 
to 2 s per slit position. Ten adjacent scanning positions,
separated by 0$\farcs$75 in the direction perpendicular to the slit, were 
sequentially measured. The sequence was repeated 250 times, with a cadence 
of 22 seconds, for a total of 2500 frames. The observation spanned approximately 92 minutes.

In addition to the full-disk images provided by ChroTel, it was possible 
to follow the evolution of the prominence using
the high-resolution \Halpha\ slit-jaw images (see Figure~\ref{Fig:SJ}).
Additionally, we downloaded images of the 304~\AA\ channel provided by the Atmospheric 
Imaging Assembly \citep[AIA,][]{Lemen2012}, on board the Solar
Dynamics Observatory \citep[SDO,][]{Pesnell2012} to have a better view
of the evolution of the prominence once it moved to the disk in further days. 
These images showed that the prominence was not linked to any
active region. Therefore, it can be considered as a quiet prominence. The
spine of the prominence remained quite stable during the observations. 
Only at the end and after the observations, the left part of the prominence showed a faster 
evolution


We applied the standard data reduction to the three observed spectral channels. 
The reduction and calibration of the data cubes included standard dark and
flat-field corrections, as well as the wavelength calibration. Flat-field data
were obtained while moving the telescope around disk centre to defocus solar
granulation. The continuum of the average flat-field spectrum was rectified and 
the resulting gradient was also applied to each individual spectrum. The final 
dark and flat-field corrected spectra were expressed in units of the corresponding
continuum at disk center for all the spectral lines ($I/I_{c,dc}$). To compute the wavelength 
calibration, we compared the average flat-field spectrum of the three lines with 
the same spectral range of the FTS atlas \citep{Brault1987}. The resulting spectral 
sampling for \Halpha, \HeD3, and \CaII\ was \hbox{0.352 pm\,pixel$^{-1}$}, 
\hbox{0.325 pm\,pixel$^{-1}$}, and \hbox{0.644 pm\,pixel$^{-1}$}, respectively.

\begin{figure}[t]
  \includegraphics[width=0.24\textwidth]{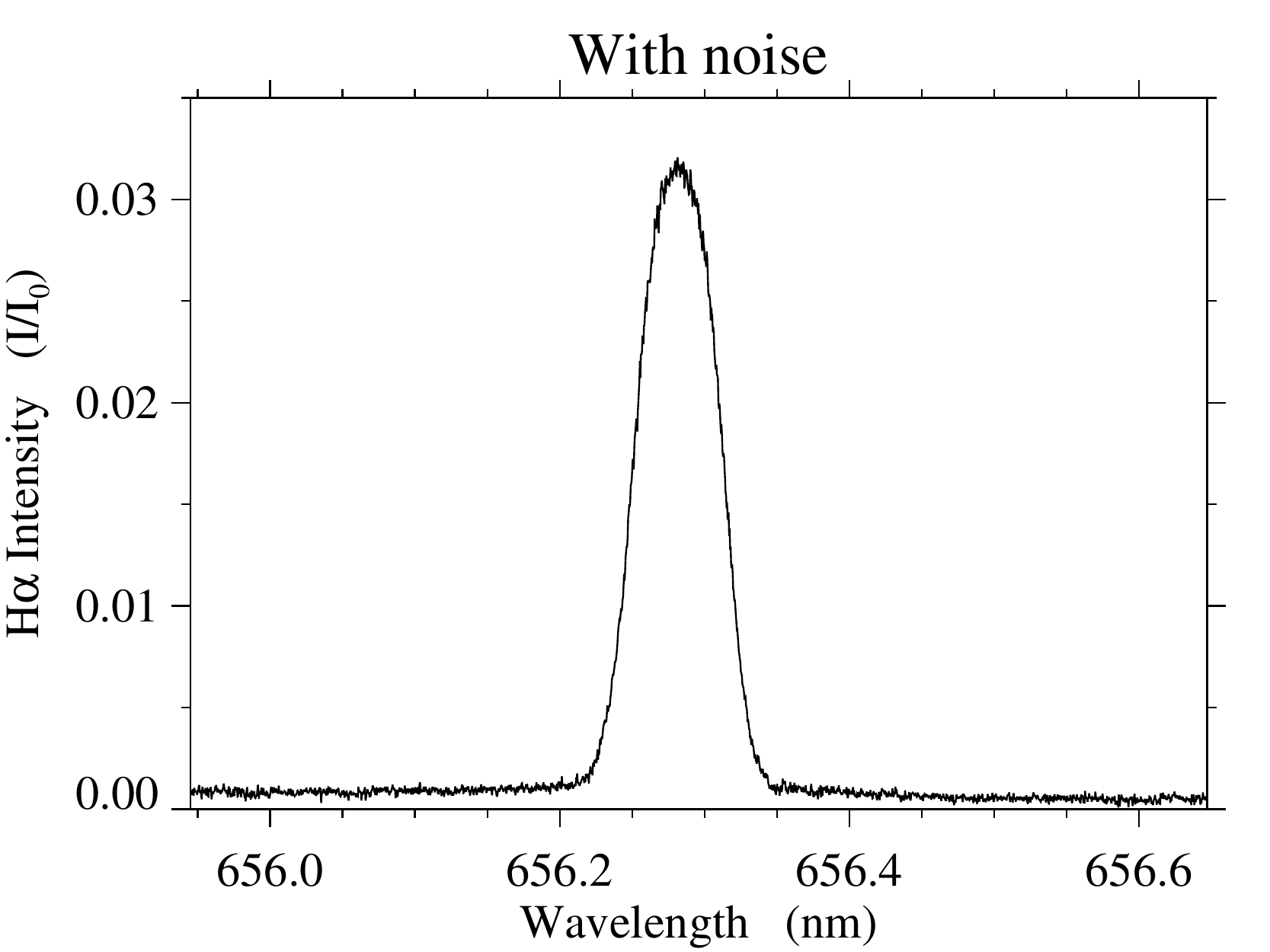}
  \includegraphics[width=0.24\textwidth]{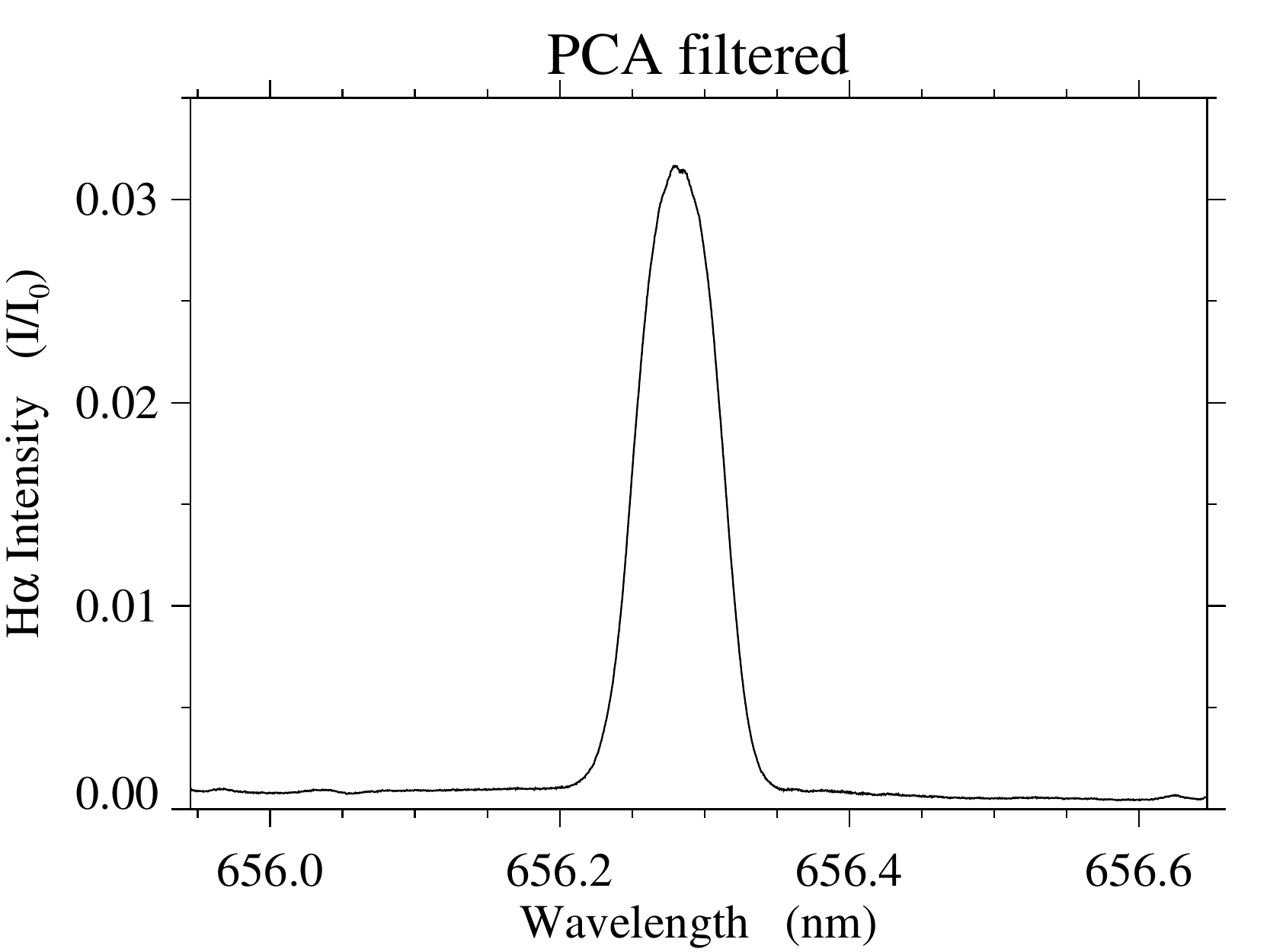}
  \caption{We present a comparison of random $\Halpha$ profiles. Random $\Halpha$ profile with noise (\emph{Left}), and random $\Halpha$ 
  profile filtered with the PCA technique {\bf(\emph{Right})}. It is clear that we reduced significantly the noise.
} 
  \label{Fig:noise}
\end{figure}

To enhance the spectral signal-to-noise ratio in our dataset, we applied a Principal Component Analysis (PCA) technique as described in \citet{Rees2000}. This approach effectively reduced noise levels. However, it's important to note that our emission profiles lack continuum normalization to unity, rendering the classical signal-to-noise ratio and its standard deviation calculation unfeasible. In Figure~\ref{Fig:noise}, we present a comparison of a random $\Halpha$ profile before and after PCA filtering, visually illustrating the noise reduction achieved by the PCA technique.
To perform the filtering using the PCA technique, each individual spectrum, $S^j(x,y,\lambda)$, 
of the three spectral channels (\Halpha, \HeD3, and \CaII, where index $j$
indicates the spectral channel) was defined as a linear combination of a set of N 
eigenvectors $e_i^j(\lambda)\ (i=1, ..., N)$  with coefficients $c_i^j(x,y)$, 
\citep[similar procedure as in][]{Khomenko2016}:
\begin{align} \label{eq:1}
    S^j(x,y,\lambda) = \sum_{i=1}^{N}\ c_i^j(x,y) e_i^j(\lambda)~. 
\end{align}
The sets of eigenvectors were calculated using 3000 randomly selected profiles 
for each spectral line by applying a singular value decomposition method 
\citep[SVD, ][]{Rees2000, Socas-Navarro2001, Dineva2020}. 
As a result of this process, the majority of the eigenvectors do not have 
information related to the shape of the spectral profiles and only carry information
of the particular noise pattern of each profile.  
The truncation of the series permits to discard information related 
to the noise and to enhance the signal-to-noise ratio. In our case, 
the expansion was truncated after the first 10 terms.

\section{Data analysis}\label{Sect:data analysis}

\subsection{Selection criteria}\label{Subsect:k-means}

The observations revealed a broad diversity of intensity profiles with different 
shapes. This variety can be related to the inhomogeneous physical conditions or dynamics 
within the prominence or the insufficient spatial resolution achieved during the observations.
In a non-negligible number of cases, the calculation of the Doppler line-of-sight (LOS)
velocity from the intensity profiles gave rise to difficulties. In this study, 
we decided to exclude profiles with a clear asymmetric or double-peaked shape. 
For the identification of the acceptable and non-acceptable profiles, we used a {\it k}-means 
clustering, an unsupervised machine learning algorithm. A similar approach is usually followed 
in previous works to gather together and classify spectral profiles in extensive data sets 
\citep[see, e.g., ][]{pietarila07, viticchie11,panos18, robustini19, 
sainz_dalda2019, Kuckein2020}. In particular, we followed the same procedure as in \citet{Kuckein2020}, 
using the {\it k}-means algorithm included in the scikit-learn library for python. 
Analysing the diversity of clusters for each spectral line, all the pixels with a clearly asymmetric or multiple-peaked shape in any of the three spectral regions were excluded.

\begin{figure}[t]
  \includegraphics[width=0.24\textwidth]{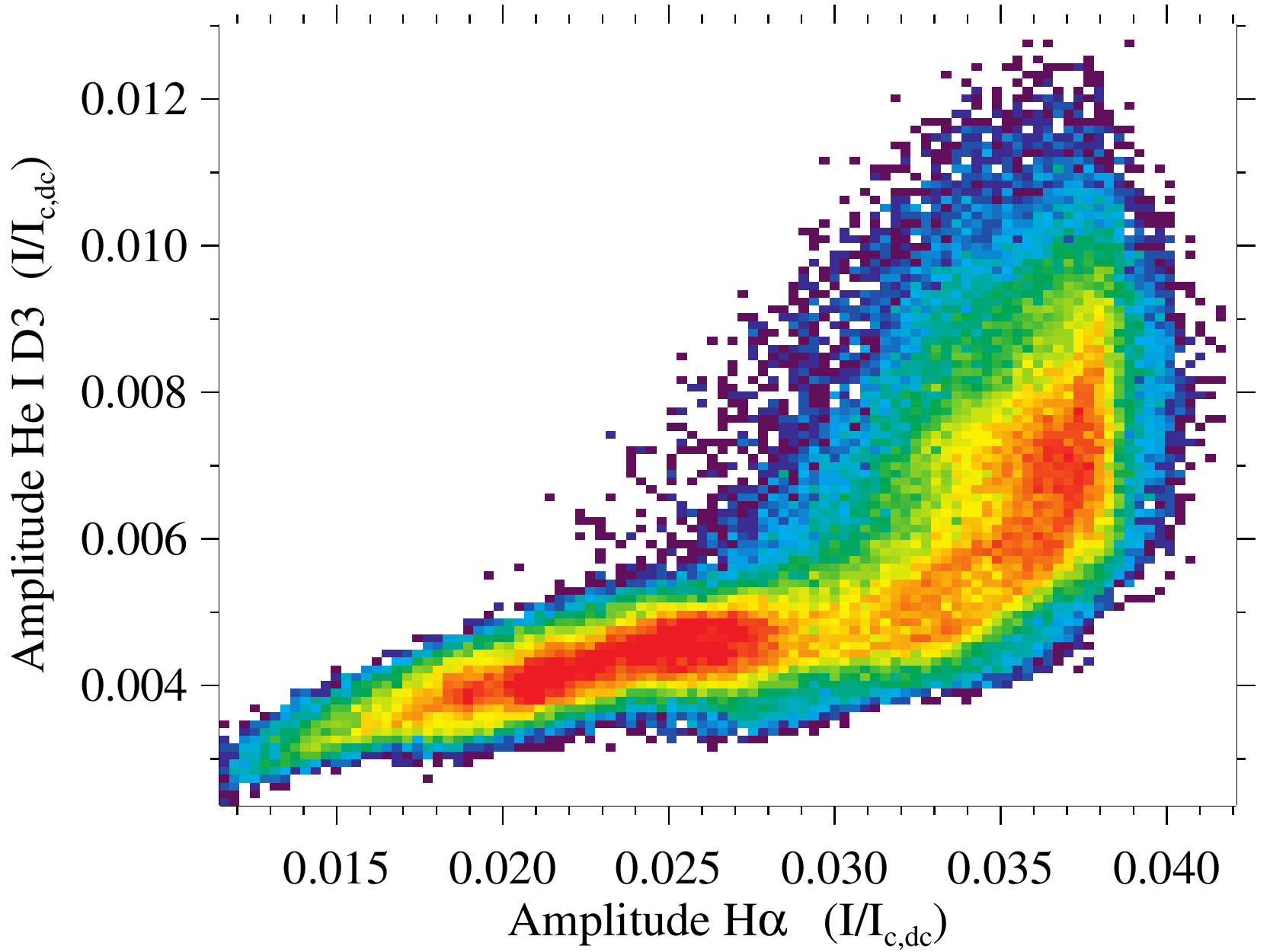}
  \includegraphics[width=0.24\textwidth]{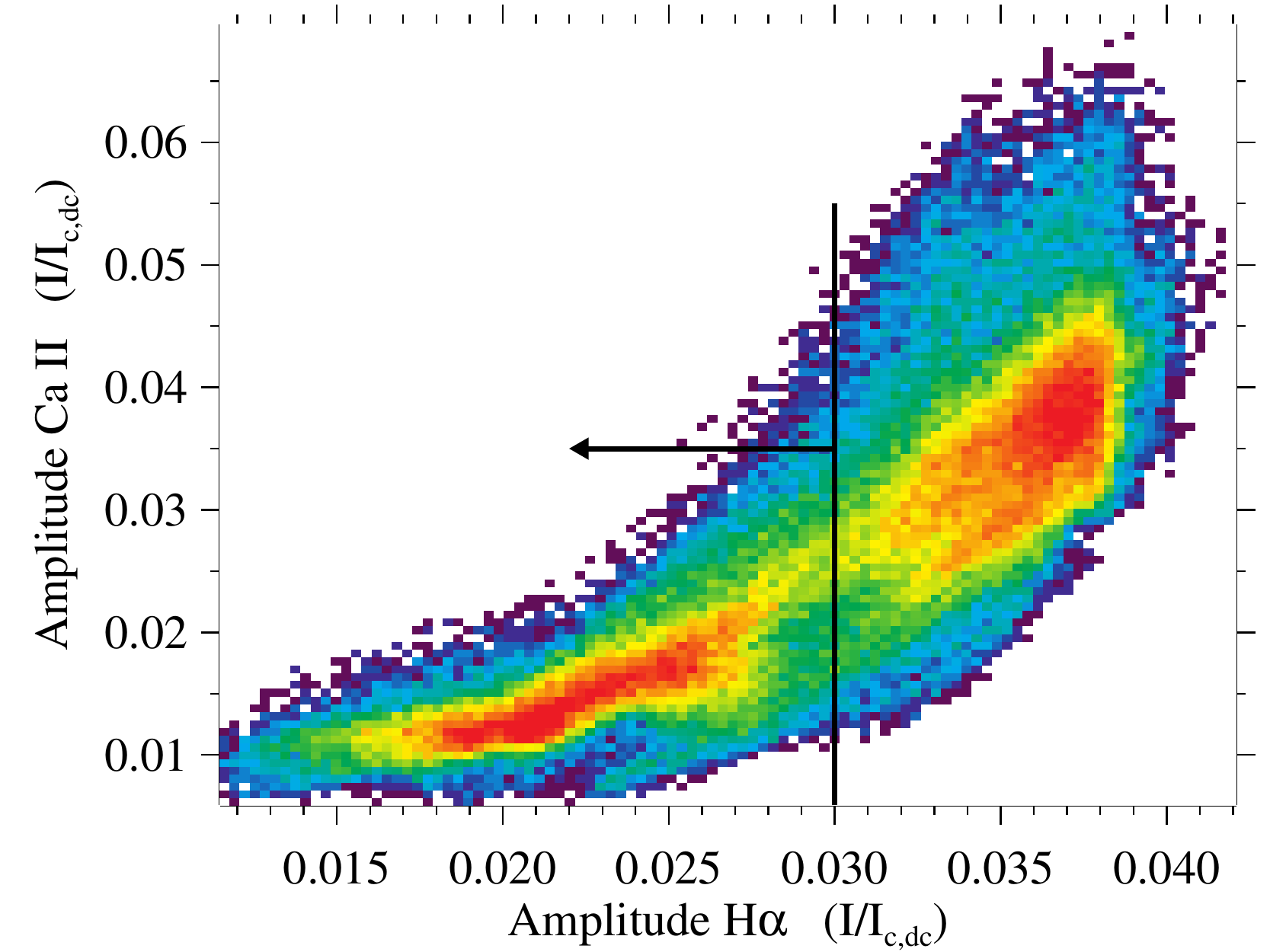}
  \caption{Two separate two-dimensional histograms illustrating dependencies between the amplitudes of different spectral lines. The dependence between the amplitudes of \Halpha\ and \HeD3\ {\bf is shown in the (\emph{Left}) panel}, whereas the Right panel depicts the dependence between the amplitudes of \Halpha\ and \CaII\. All amplitudes are normalised to the corresponding local continuum at disk centre. Redder/bluer colors mean a larger/smaller density of points. A linear relationship exists in both plots for \Halpha\ normalised amplitudes below 0.03. The arrow indicates the interval considered as optically thin for the analysis (see Sect.\,\ref{Subsect:k-means}.)
} 
  \label{Fig:amplitudes}
\end{figure}

\begin{figure}[!t]
    \centering
    \begin{subfigure}{0.16\textwidth}
    \includegraphics[width=1\textwidth]{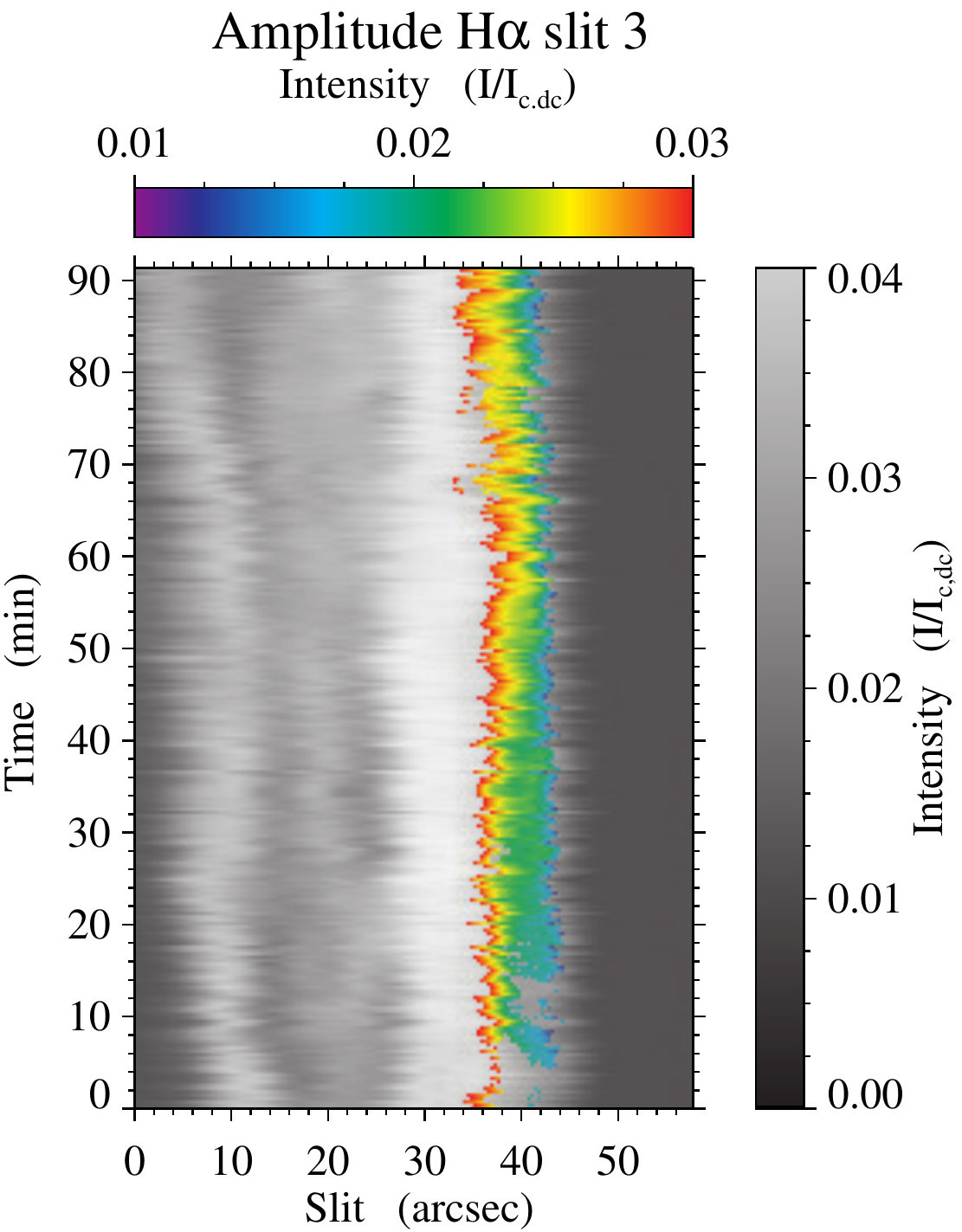}
    \end{subfigure}
    \begin{subfigure}{0.16\textwidth}
    \includegraphics[width=1\textwidth]{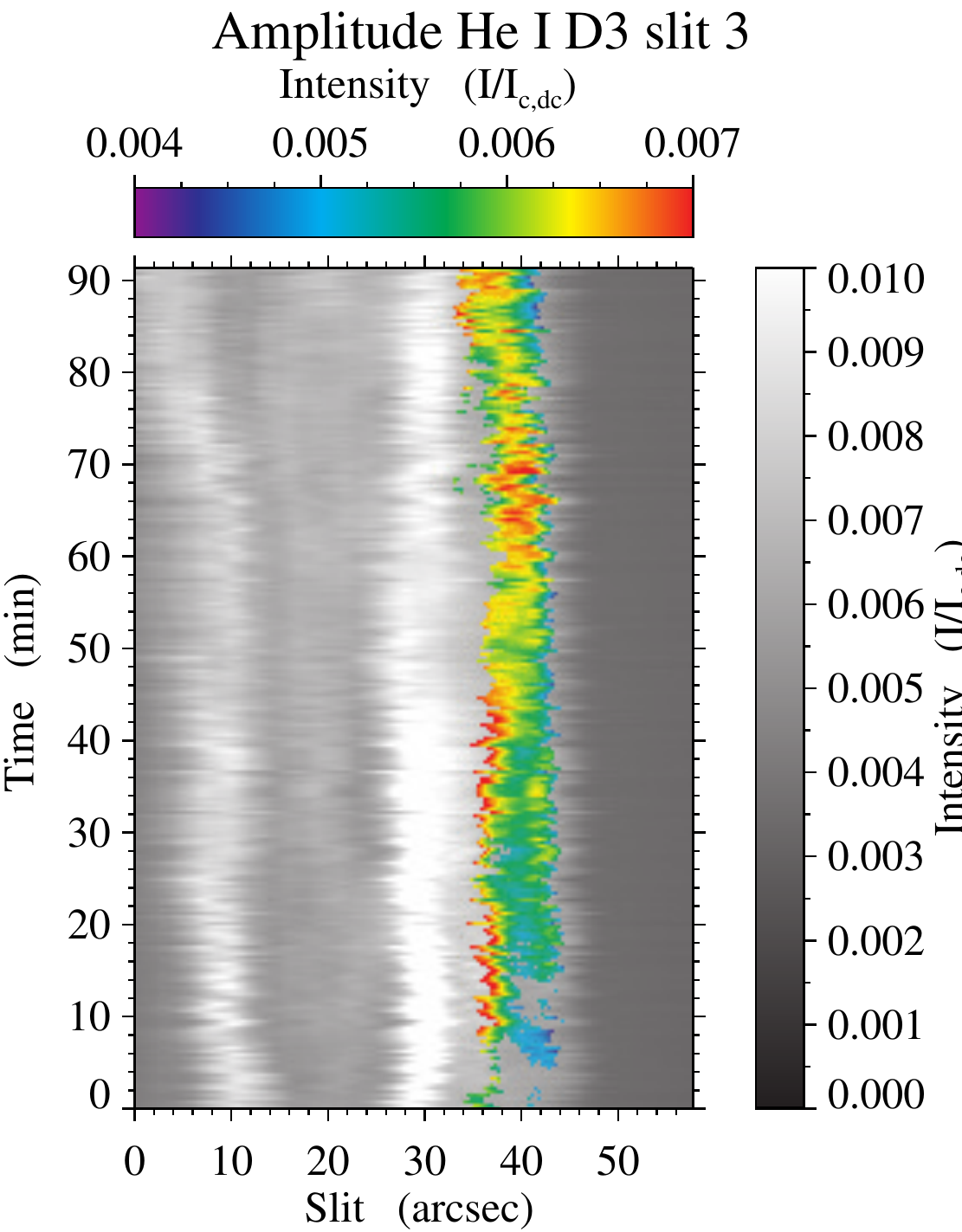}
    \end{subfigure}
    \begin{subfigure}{0.16\textwidth}
    \includegraphics[width=1\textwidth]{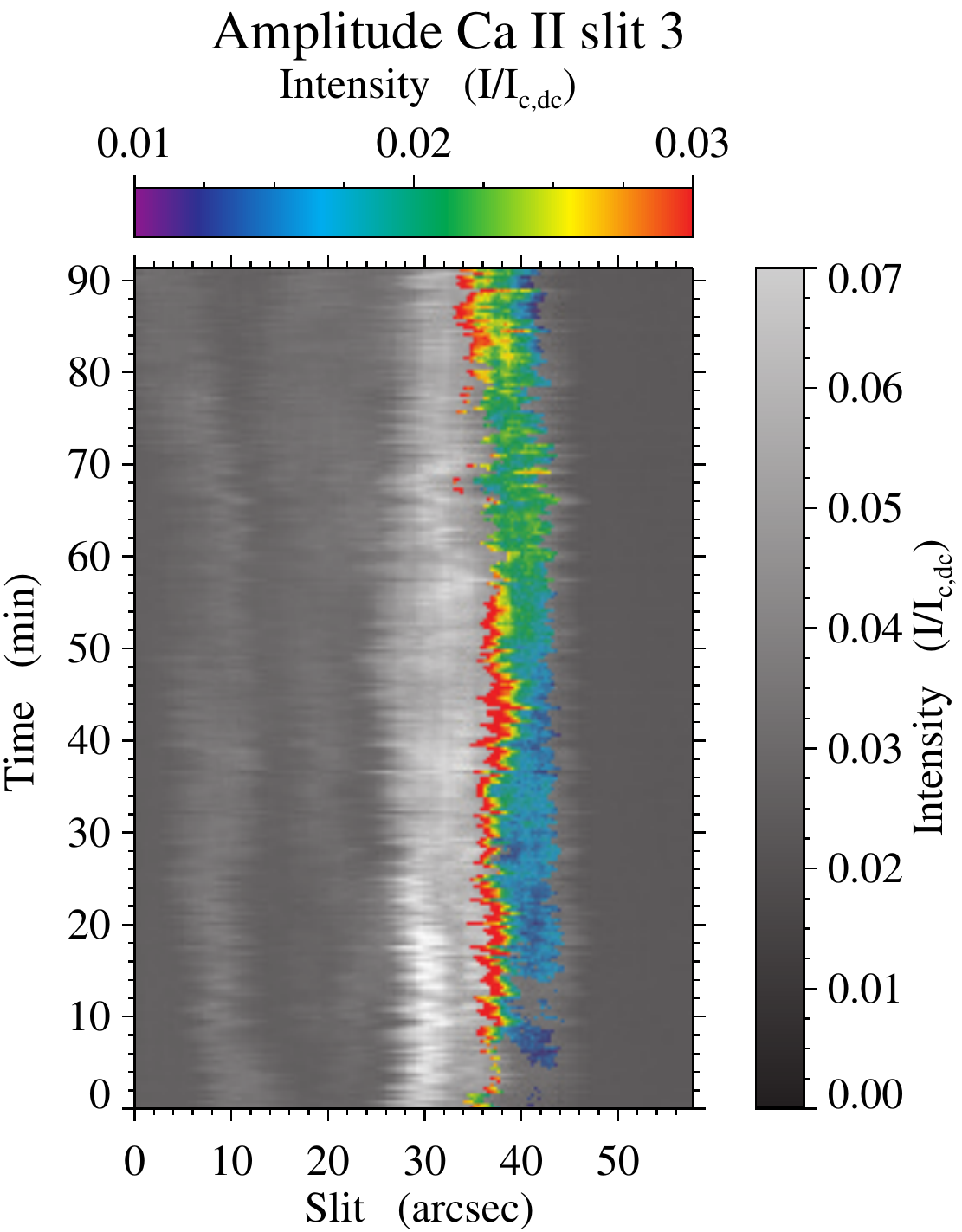}
    \end{subfigure}\\[1ex]
    \begin{subfigure}{0.16\textwidth}
    \includegraphics[width=1\textwidth]{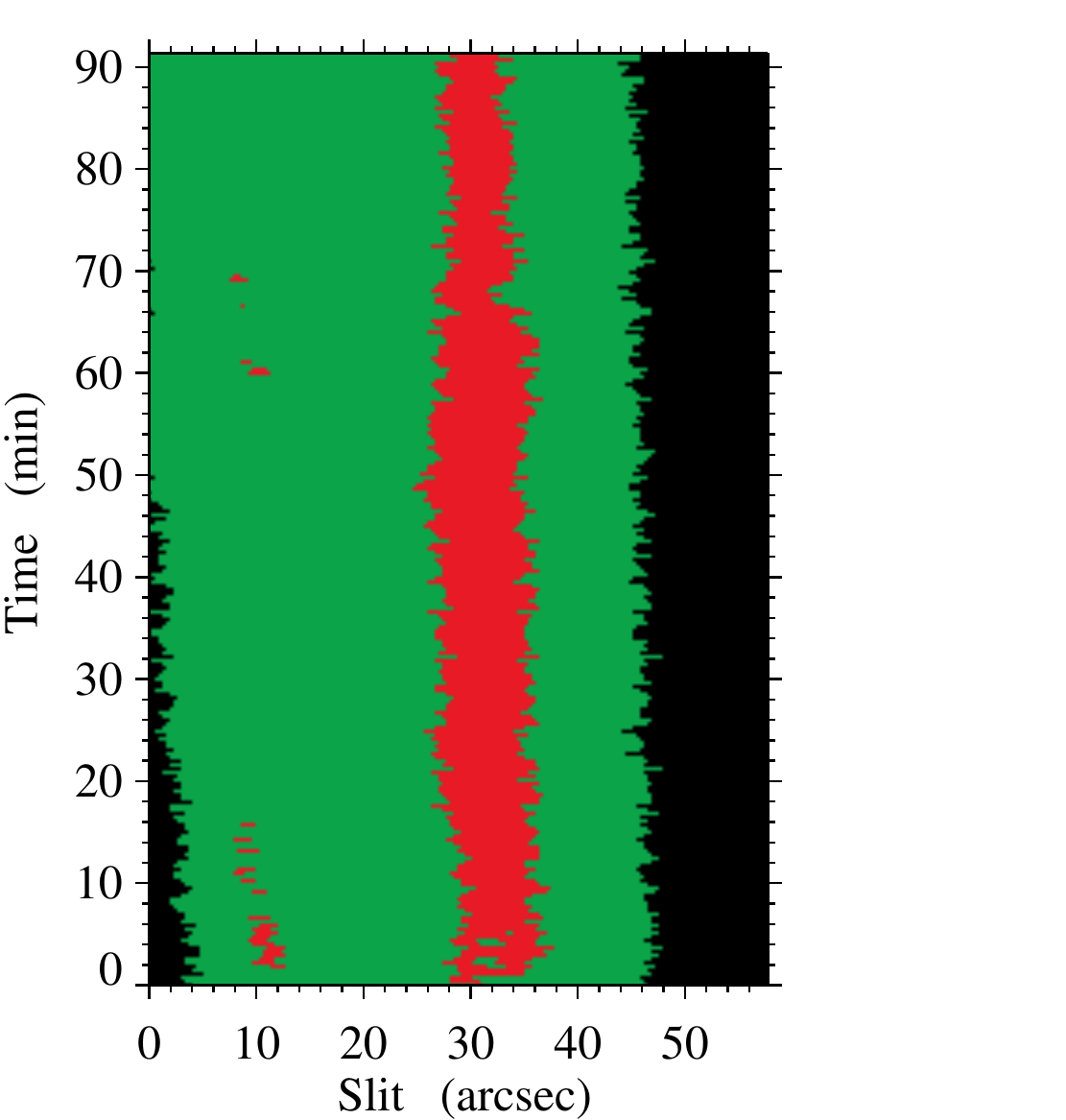}
    \end{subfigure}
    \begin{subfigure}{0.16\textwidth}
    \includegraphics[width=1\textwidth]{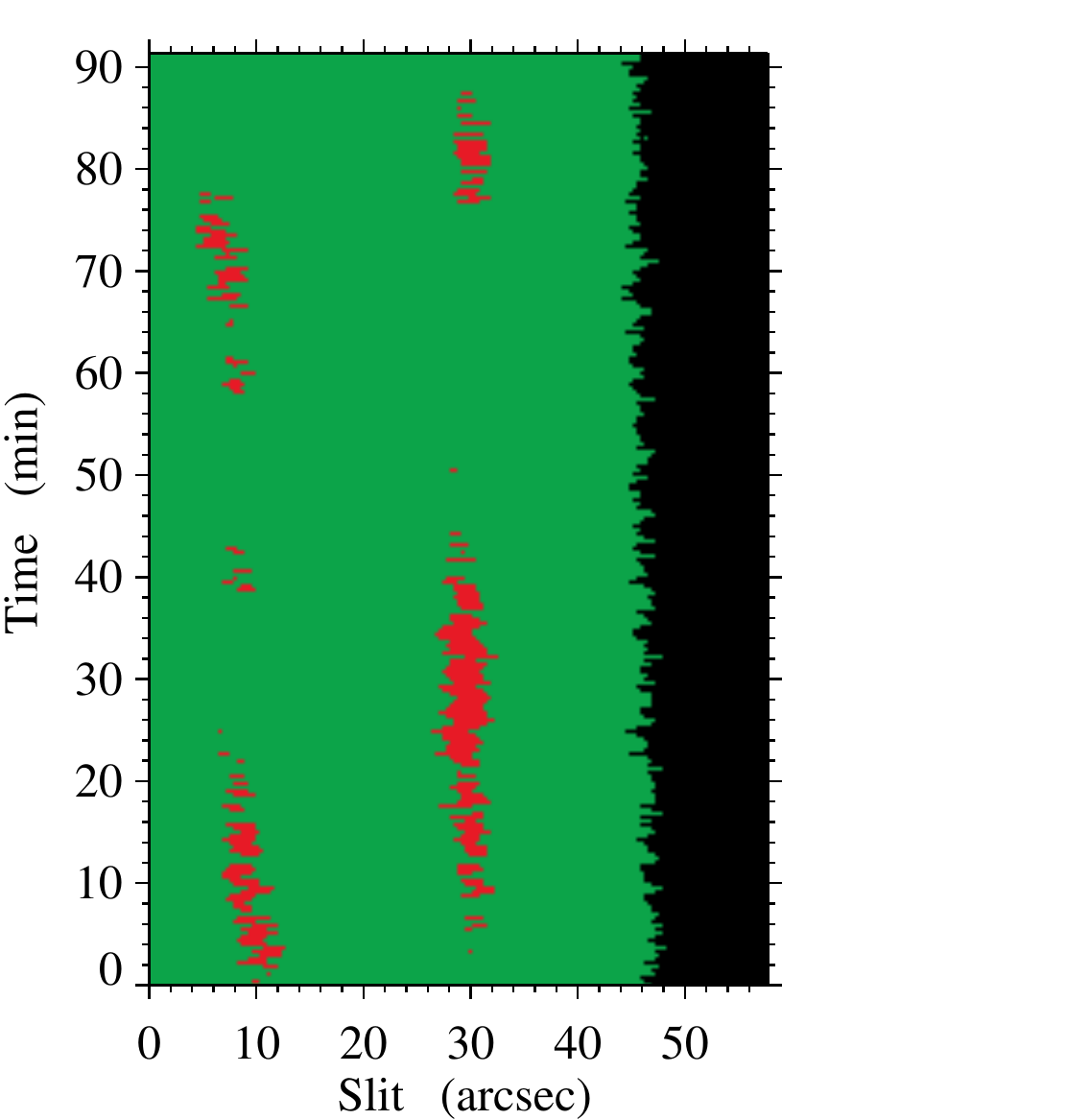}
    \end{subfigure}
    \begin{subfigure}{0.16\textwidth}
    \includegraphics[width=1\textwidth]{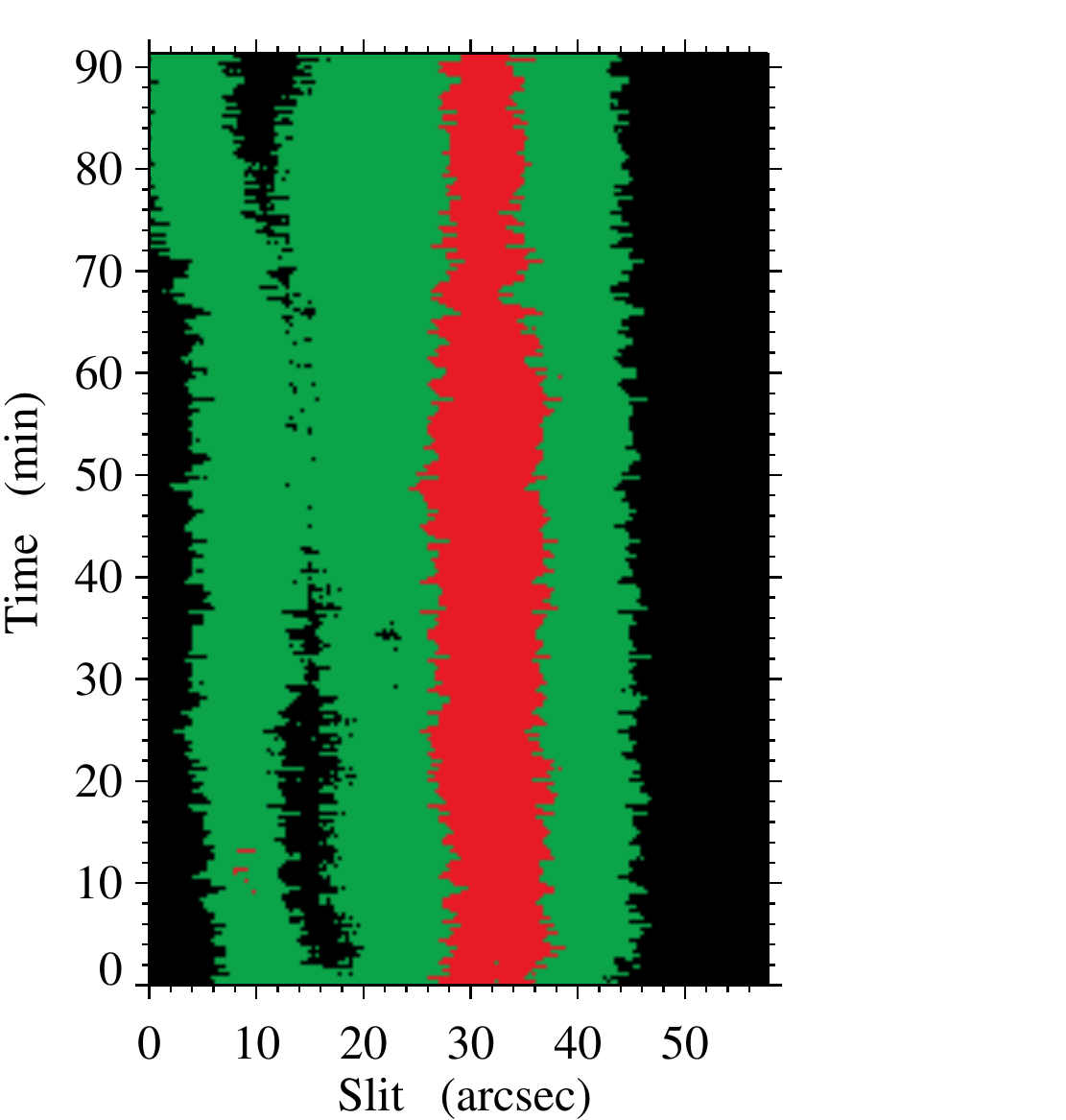}
    \end{subfigure}
    \caption{In this figure, we represent the time-slit intensity maps and the optically thin criteria for different spectral lines. \emph{Top:} In grey, time-slit maps of the maximum intensity in the three observed spectral lines for the scan position number 3, \Halpha\ (left), \HeD3\ (centre), \CaII\ (right). In colour, the optically thin pixels with regular symmetric profiles used for the analysis. \emph{Bottom:} same time-slit maps with the criterion by \citet{Zapior2022} to define points that are optically thin (green) or thick (red). The black region represent the very small amplitudes rejected for the analysis.
    }  
    \label{Fig:amplitude_maps}
\end{figure}

Complementing the above criterion, we have also restricted the analysis to those locations
where the plasma is optically thin \citep[similar approach as in][]{Khomenko2016}.
To find the regions where the plasma can be considered as optically thin, we compared the \Halpha\ 
maximum intensity amplitude (the intensity peak) with the same parameter of the other two spectral lines 
(see Fig.\,\ref{Fig:amplitudes}). 
The relationship moves away from the linearity approximately when the amplitude of \Halpha\  
exceeds a certain value ($I/I_0 \sim 0.03$, where $I_0$ is the continuum intensity at disk centre), 
indicating that \Halpha\ enters a saturation regime well before the other lines, which remain less intense. 
Therefore, we also excluded all pixels with \Halpha\ amplitudes larger than this threshold. 
In Figure~\ref{Fig:amplitude_maps} (upper figures) we represent the amplitude
in a fixed slit position (position index 3 of the spectrograph slit) for the three spectral lines.
The gray zone is the excluded zone, either because it contains asymmetric or multiple-peaked profiles 
or is considered optically thick. The coloured part represents the available pixels for the subsequent
analysis. The region that fits our selection criteria is at the edge of the prominence.

\begin{figure*}
    \centering
    \begin{subfigure}{0.33\textwidth}
    \includegraphics[width=1\textwidth]{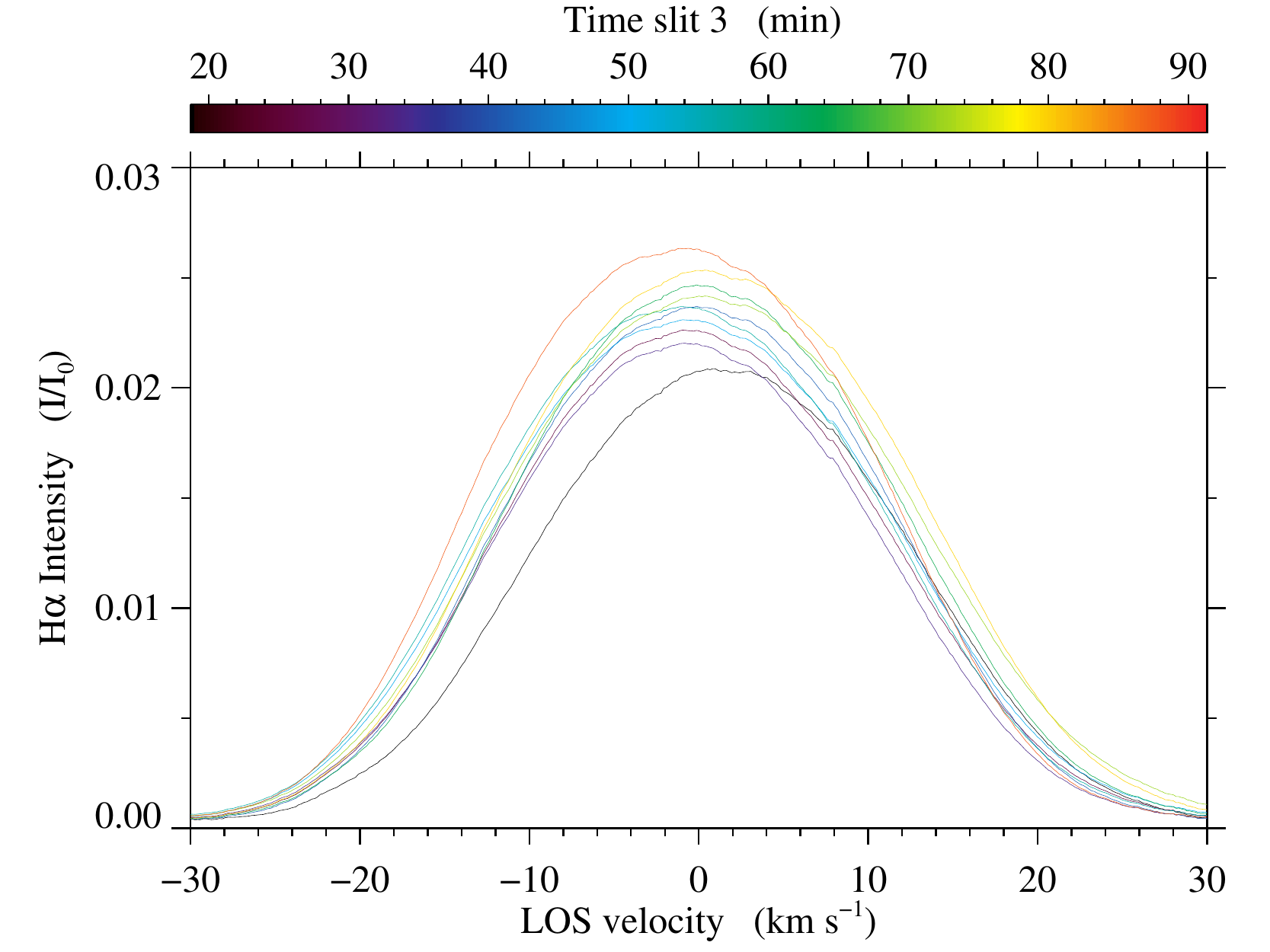}
    \end{subfigure}
    \begin{subfigure}{0.33\textwidth}
    \includegraphics[width=1\textwidth]{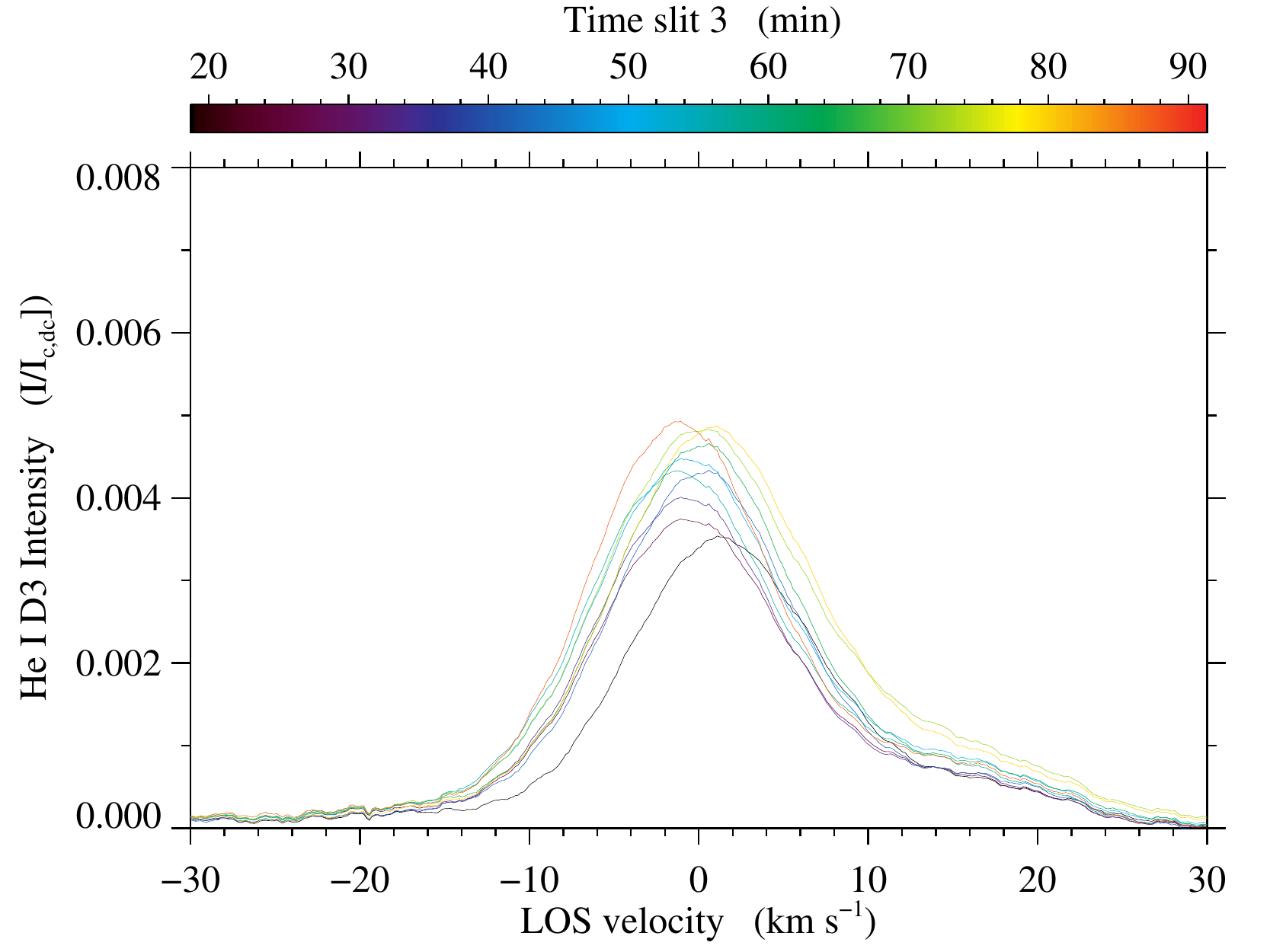}
    \end{subfigure}
    \begin{subfigure}{0.33\textwidth}
    \includegraphics[width=1\textwidth]{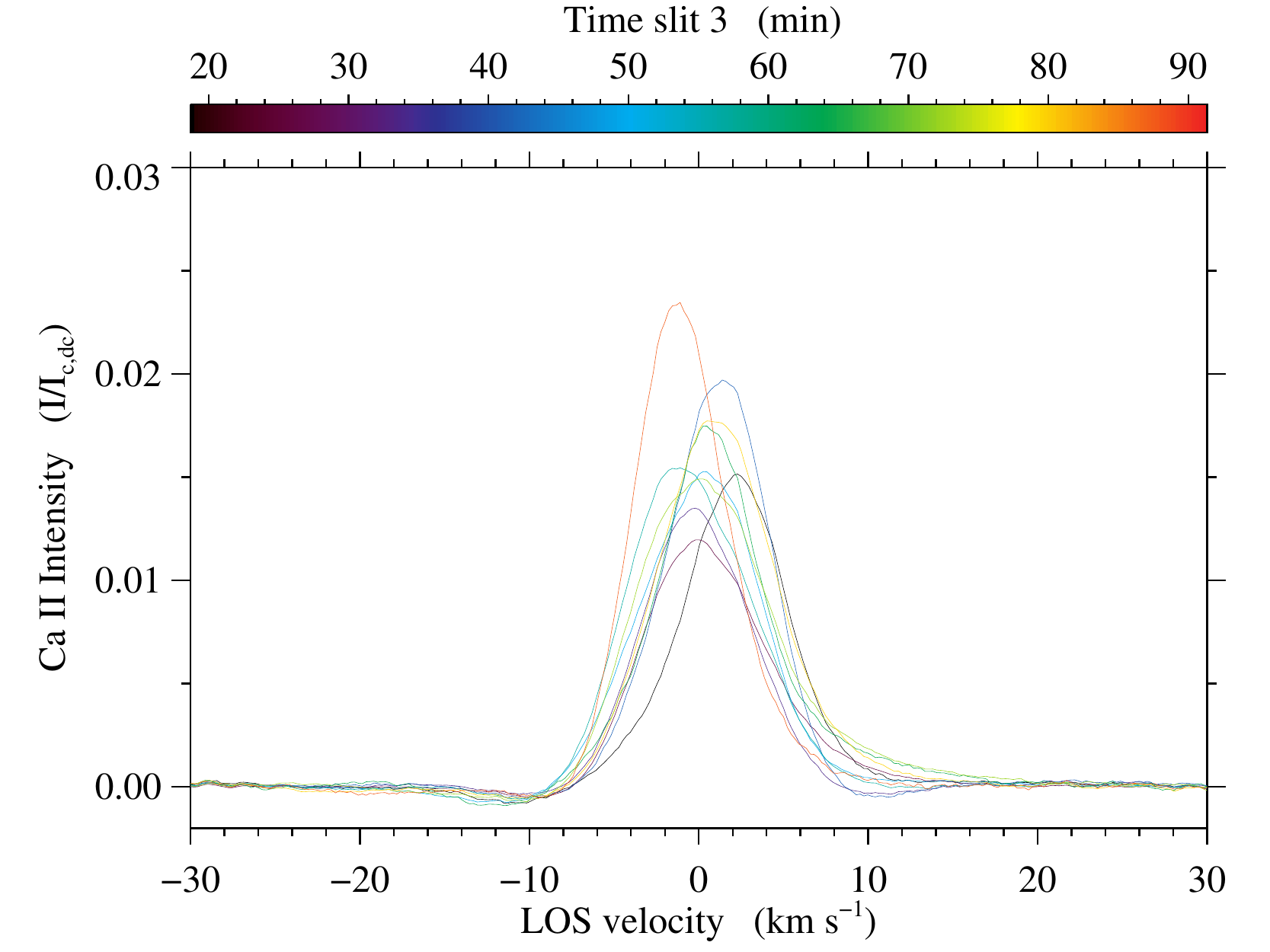}
    \end{subfigure}
    \caption{Randomly selected spectral profiles that fulfill the criteria of being optically thin, symmetric and single-peaked (see Sect.\,\ref{Subsect:k-means}). \emph{Left}: \Halpha\ profiles, \emph{Centre}: \HeD3\ profiles (note the unavoidable asymmetry introduced by the red component), and \emph{Right}: \CaII\ profiles.
    }
    \label{Fig:profiles}
\end{figure*}

In order to ensure that only pixels corresponding to optically thin spectral lines were considered, we also compared our criterion for optical thickness with that presented in \citet{Zapior2022}, based on the total energy ($E_\mathrm{tot}$) emitted by each spectral line.
To determine the line-center optical thickness ($\tau$), we integrated the (normalised) spectral line profiles and multiplied them by the local continuum intensity at disk centre. The continuum intensity values were obtained from the Harvard-Smithsonian reference atmosphere  \citep[HSRA]{Gingerich}, which provides reliable and accurate measurements of the solar continuum at the observed wavelengths.
To define the optically thin regime, we used the criterion $\tau \leq 0.5$. Based on the calculations presented by \citet{Zapior2022}, for \Halpha\ to meet this criterion, $E_\mathrm{tot}$ should be less than $6 \times 10^4$ erg s$^{-1}$ cm$^{-2}$ sr$^{-1}$, for \HeD3, $E_\mathrm{tot}$ should be less than $3 \times 10^5$ erg s$^{-1}$ cm$^{-2}$ sr$^{-1}$, and for \CaII, $E_\mathrm{tot}$ should be less than $1 \times 10^4$ erg s$^{-1}$ cm$^{-2}$ sr$^{-1}$. The lower part of Figure~\ref{Fig:amplitude_maps} illustrates the optically thin regime in green \hbox{($\tau \leq 0.5$)} for each spectral line, using the index 3 slit map. Regions in red ($\tau > 0.5$) correspond to the optically thick regime, while black regions indicate areas with very small amplitudes that were rejected for the analysis. As can be seen, our criterion based on the saturation of the \Halpha\ amplitude, complemented with the condition of having symmetric single-lobed profiles, is more restrictive than the one based on the total energy emitted by the spectral lines. 
We preferred to keep our conditions as conservative criteria to ensure that 
the velocities for the three spectral lines provide accurate and reliable comparisons in the same plasma volume, and are not affected by significant LOS velocity gradients or horizontal LOS velocity variations within the spatial size of our pixels. 
Figure\,\ref{Fig:profiles} provides some examples of the retained spectral profiles. As expected, all spectral lines exhibit a regular shape that is consistent with our predefined selection criteria. The most noticeable variation among the different profiles is the presence of red or blueshifted features. 

\begin{figure}[!t]
    \centering
    \begin{subfigure}{0.16\textwidth}
    \includegraphics[width=1\textwidth]{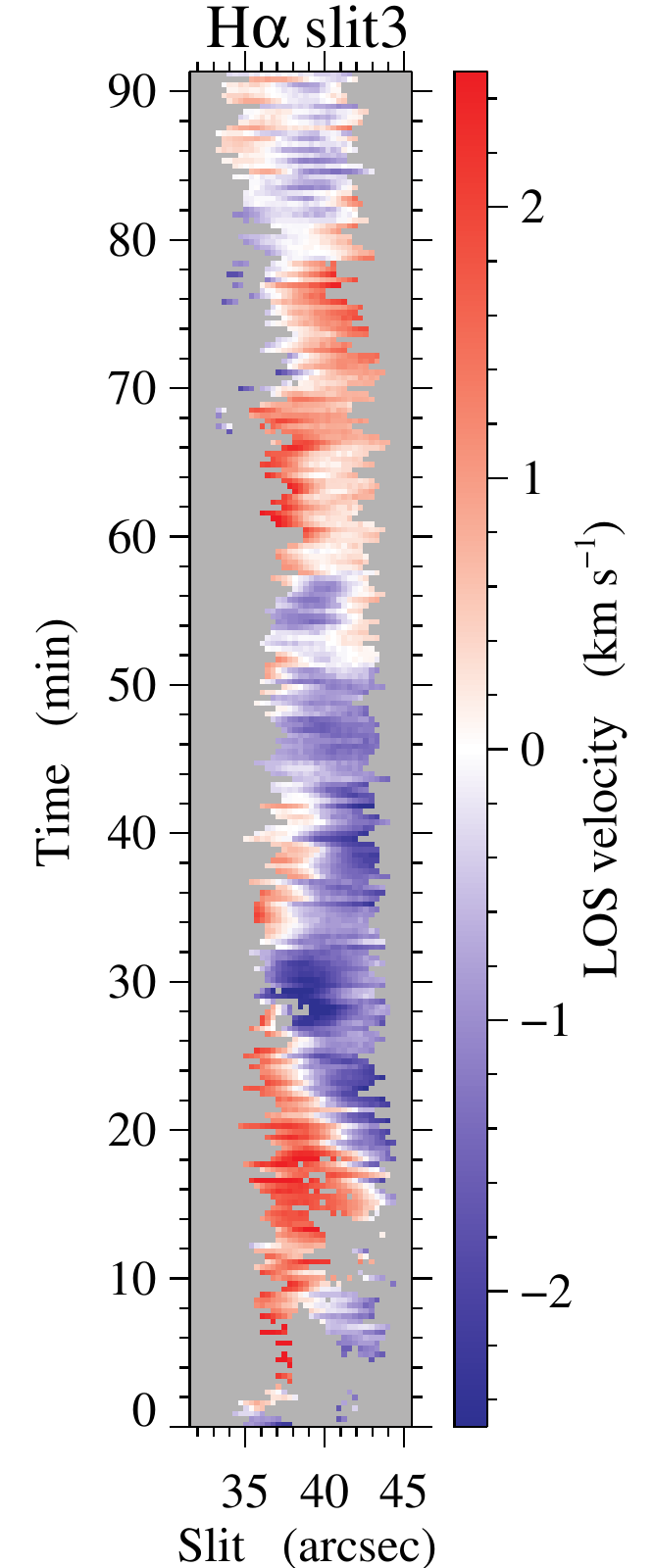}
    \end{subfigure}
    \begin{subfigure}{0.16\textwidth}
    \includegraphics[width=1\textwidth]{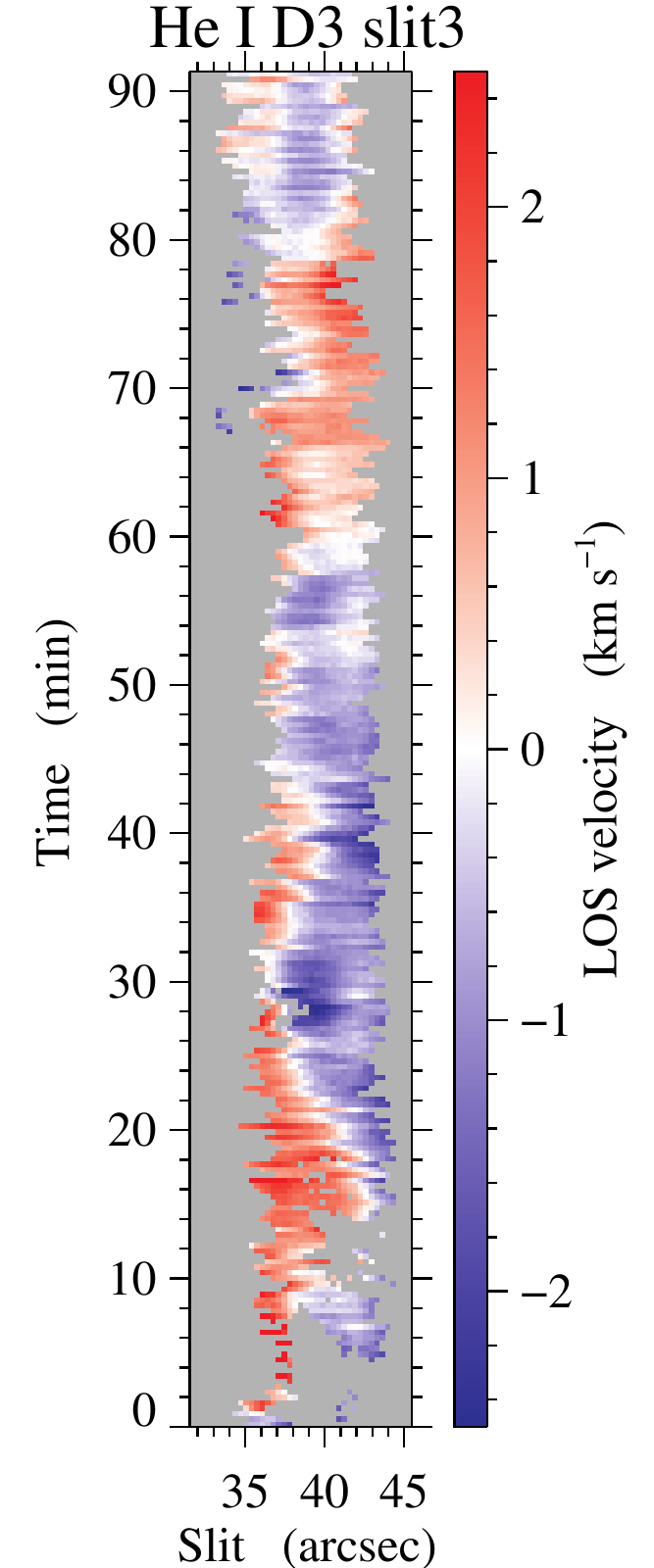}
    \end{subfigure}
    \begin{subfigure}{0.16\textwidth}
    \includegraphics[width=1\textwidth]{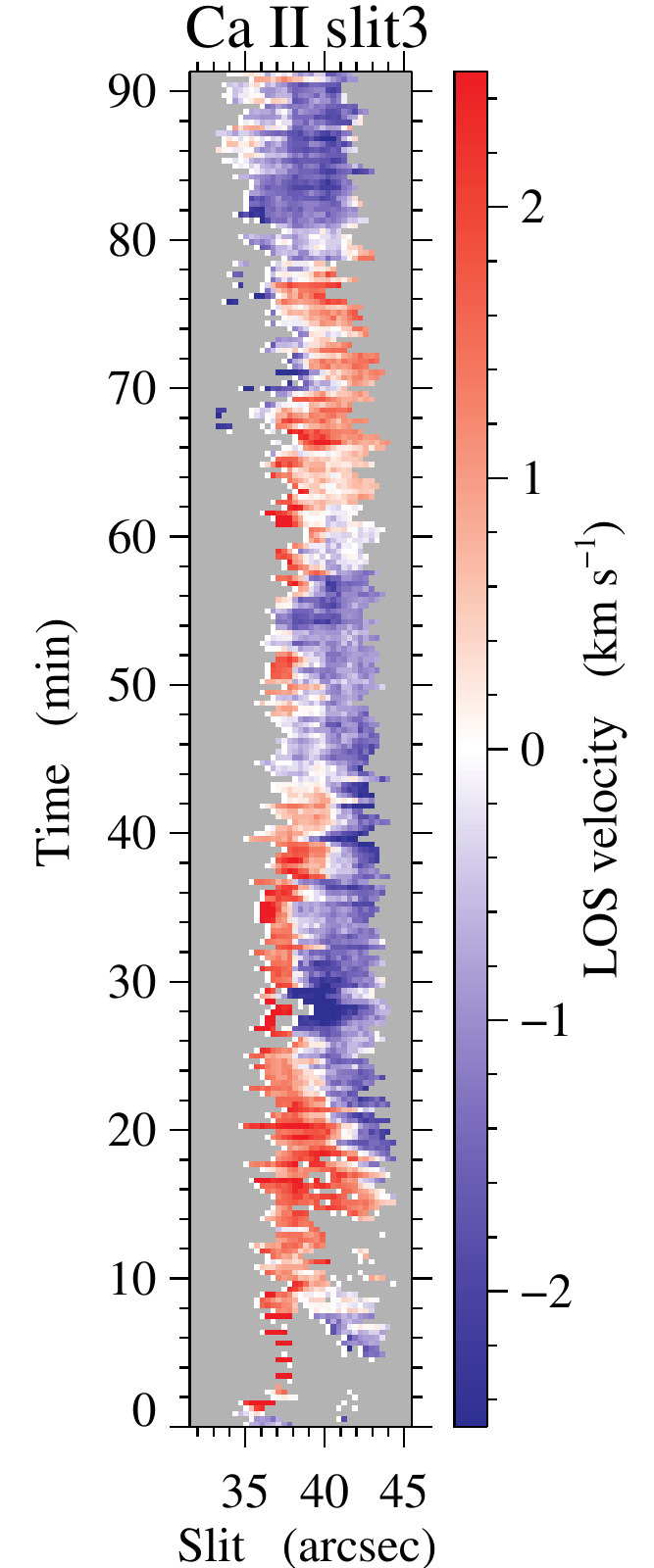}
    \end{subfigure}\\[1ex]
    \begin{subfigure}{0.16\textwidth}
    \includegraphics[width=1\textwidth]{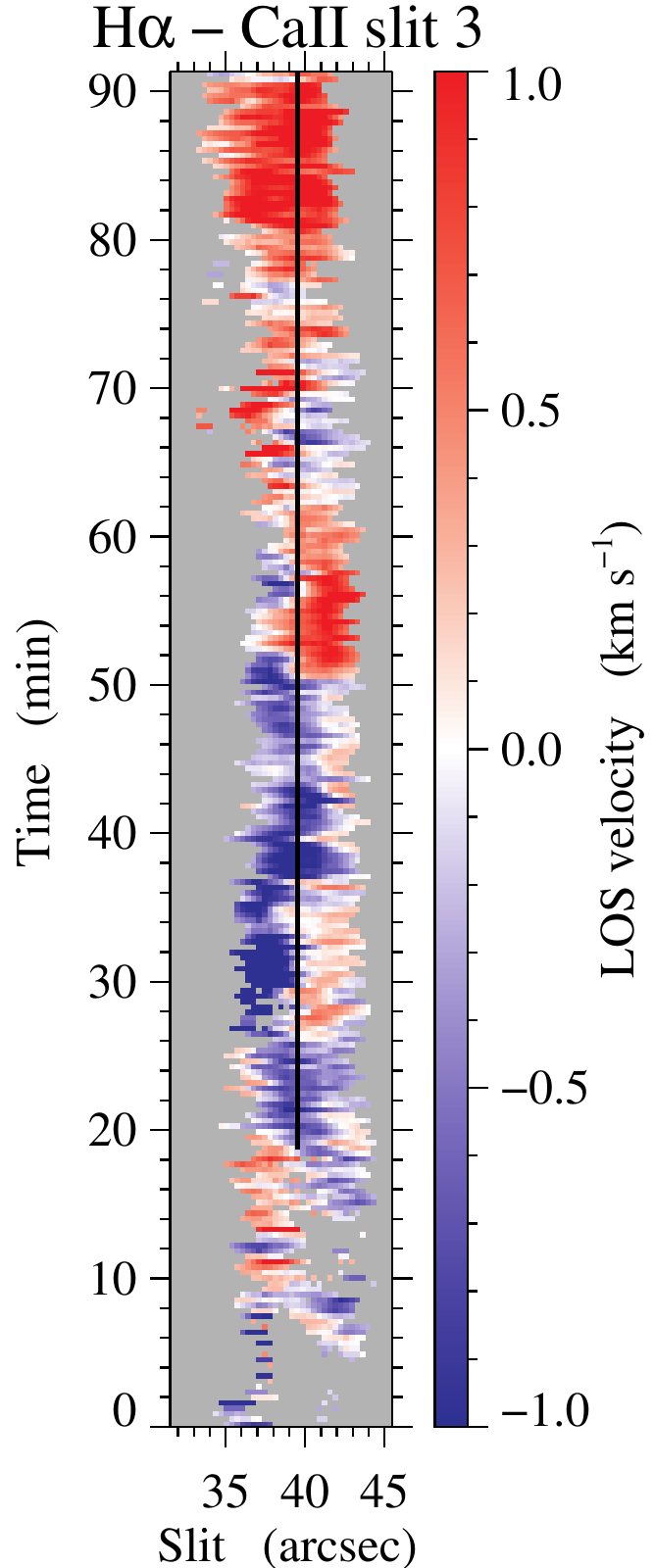}
    \end{subfigure}
    \begin{subfigure}{0.16\textwidth}
    \includegraphics[width=1\textwidth]{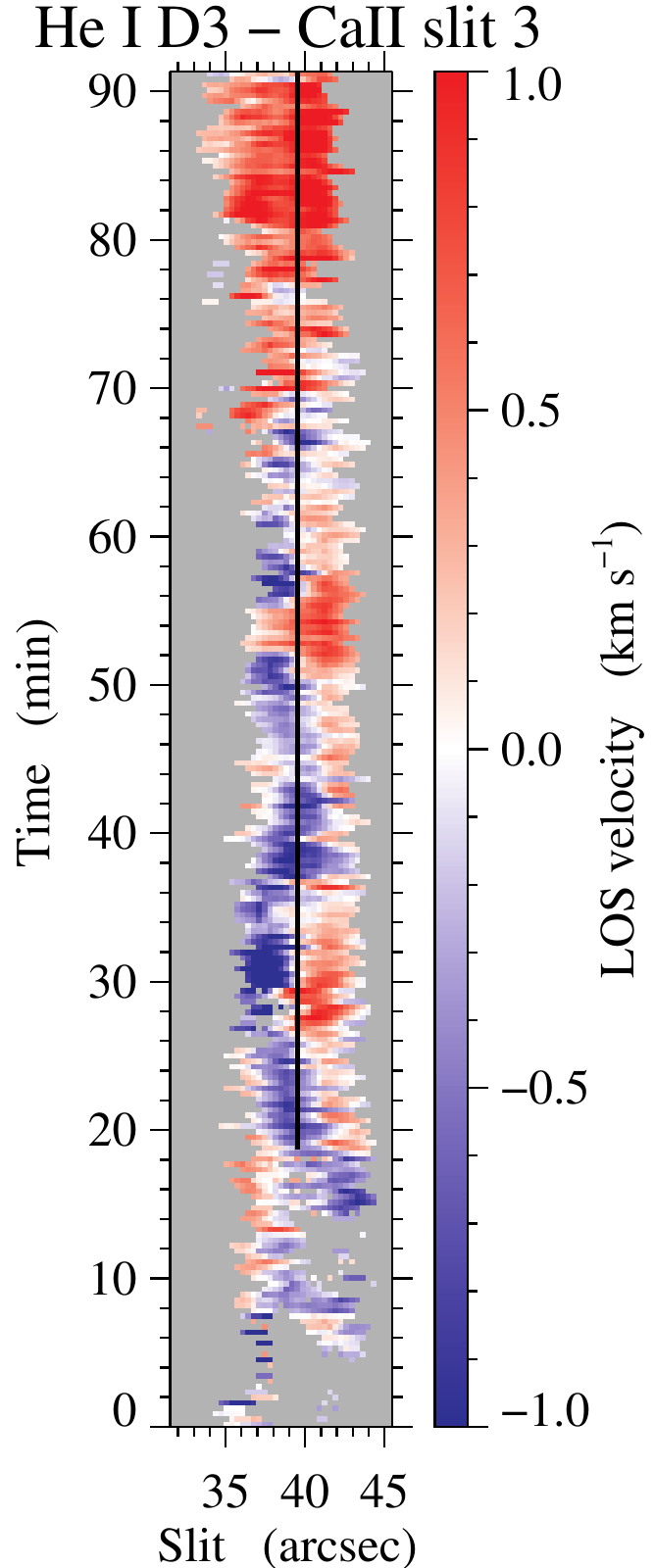}
    \end{subfigure}
    \begin{subfigure}{0.16\textwidth}
    \includegraphics[width=1\textwidth]{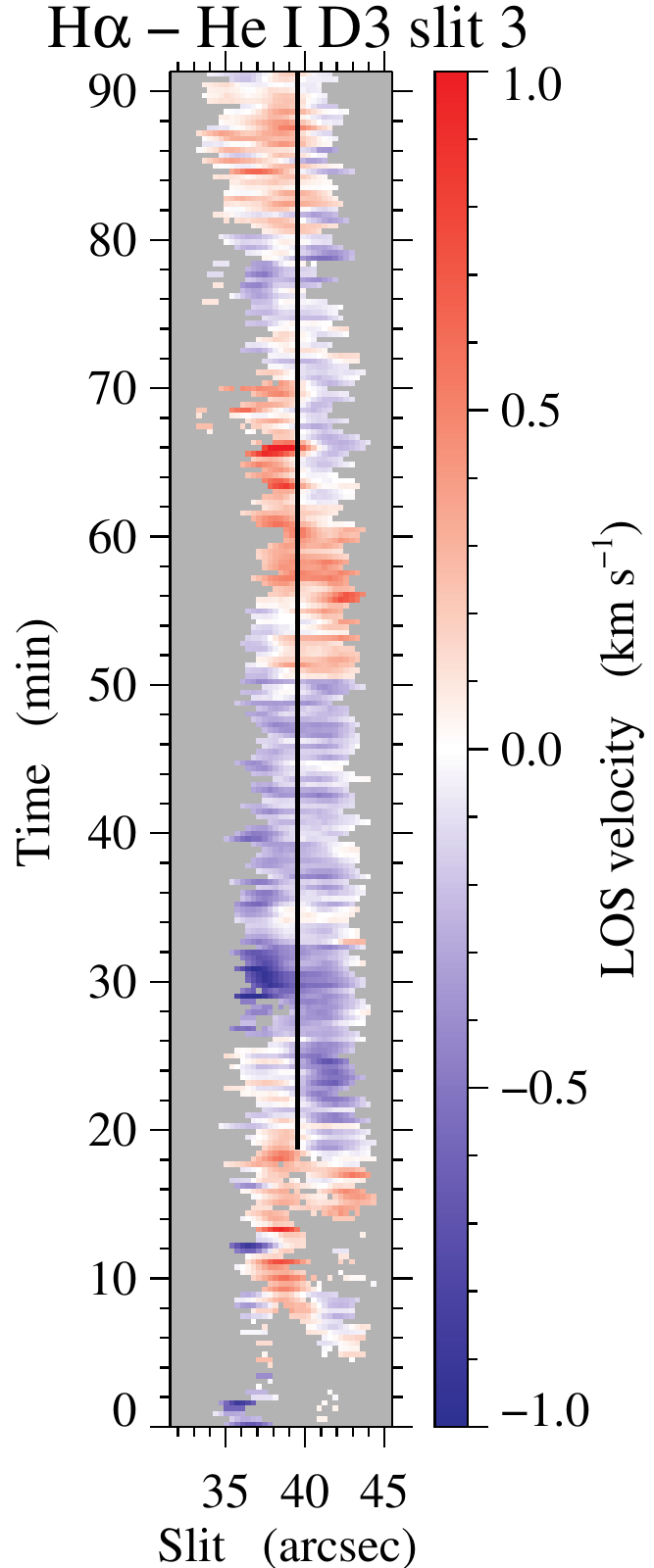}
    \end{subfigure}
    \caption{Illustration of time-slit velocity maps and velocity differences at scan position 3. \emph{Top:} Time-slit velocity maps of scan position number 3. The scale is clipped at $\pm\,2.5$ km\,s$^{-1}$ for the three spectral lines, \Halpha\ (left), \HeD3\ (centre), \CaII\ (right), for the selected points in the field of view. Velocities have been calculated applying the center-of-gravity method to the upper 20\% of each spectral line. \emph{Bottom:} Velocity differences calculated between pairs of lines as indicated in the label of each plot. In this case, the scale es clipped at $\pm 1$ km\,s$^{-1}$. The vertical black line in the bottom line is used to display the temporal variations of the velocities in Fig.\,\ref{Fig:v_variation}. 
    }
    \label{Fig:Doppler}
\end{figure}

\subsection{LOS Velocity determination}\label{Subsect:fitting}

The shape of the selected spectra of the \Halpha\ and \CaII\ spectral lines consists of a single emission line profile (\Halpha\ is indeed composed by a number of transitions between levels $n=3$ and $n=2$, but they are all so close together in energy that they can be considered as a single transition for our purposes). On the other hand, the \HeD3\ multiplet exhibits a double emission line profile. This multiplet is composed of six transitions, where the blue component combines five emission transitions and the weaker red component is produced by the sixth transition \citep{lopez-ariste2015,Koza2017}.

In order to determine the most reliable method for obtaining the line-of-sight (LOS) velocities, we applied several line-fitting methods and evaluated their performance. On the one hand, we performed a polynomial fitting to the core of the \Halpha, \CaII, and the blue component of the \HeD3\ emission lines. For the fitting process, we chose to fit the core of the spectral lines using a second-degree least-square polynomial fit. To determine the region of the line core for each spectral line, we identified the maximum intensity value of the line and selected a range of spectral intensity points at both sides of the maximum. The number of spectral points selected varied depending on the width of the spectral line.

Additionally, Gaussian profiles were employed to fit the three spectral lines. This method allowed us to characterise the full line profiles to extract the necessary information for velocity calculations.

Complementarily, we applied the center of gravity method (COG) with different intensity thresholds, specifically considering values normalised to the maximum of $I/I_{\text{max}} > 0.2$, $0.5$, and $0.8$. The COG method calculates the weighted average of the positions of the spectral intensity along the line profile. 

In general, we found that all methods generally yielded similar global results, although some local differences were observed. The COG method provided smoother LOS velocity results and demonstrated less sensitivity to spectral noise or to the global shape of the spectral lines compared to the polynomial fitting and the Gaussian profile methods. The polynomial fit to the core of the line gave wrong results in some pixels because of particular noise that led to the determination of a wrong maximum around which the fit was done. Expanding the interval for the fit was of no help, because then the line shape deviated significantly from a second order polynomial and the fit was not accurate enough. Third and fourth order polynomials were also tried, but they did not improve the results. The Gaussian fit to the whole profile was instead sensitive to slight asymmetries appearing in the wings. The COG method proved to be much less sensitive to all these issues. For all these reasons, we adopted the intensity threshold of $I/I_{\text{max}} > 0.8$ as the preferred criterion for comparing the LOS velocities between the three spectral lines. Finally, since we are only interested in the analysis of relative velocities, we subtracted the average prominence velocity at each spectral band in the selected field of view (see Sect. \ref{Subsect:k-means}). The resulting time-slit velocity maps of scan position number 3 that were obtained using the latter procedure are shown in the top panels of Fig.~\ref{Fig:Doppler}.

In all cases, the formal errors of the velocity determination were extremely small. We prefer to follow a conservative approach and assign an uncertainty equivalent to $\pm\,0.5$ pixels in each spectral range. According to the dispersion, this criterion is equivalent to a velocity uncertainty of \hbox{$\pm\,80$ \ms} in \Halpha, \hbox{$\pm\,83$ \ms} in \HeD3 and \hbox{$\pm\,113$ \ms} in \CaII.

\begin{figure*}
    \centering
    \begin{subfigure}{0.33\textwidth}
    \includegraphics[width=1\textwidth]{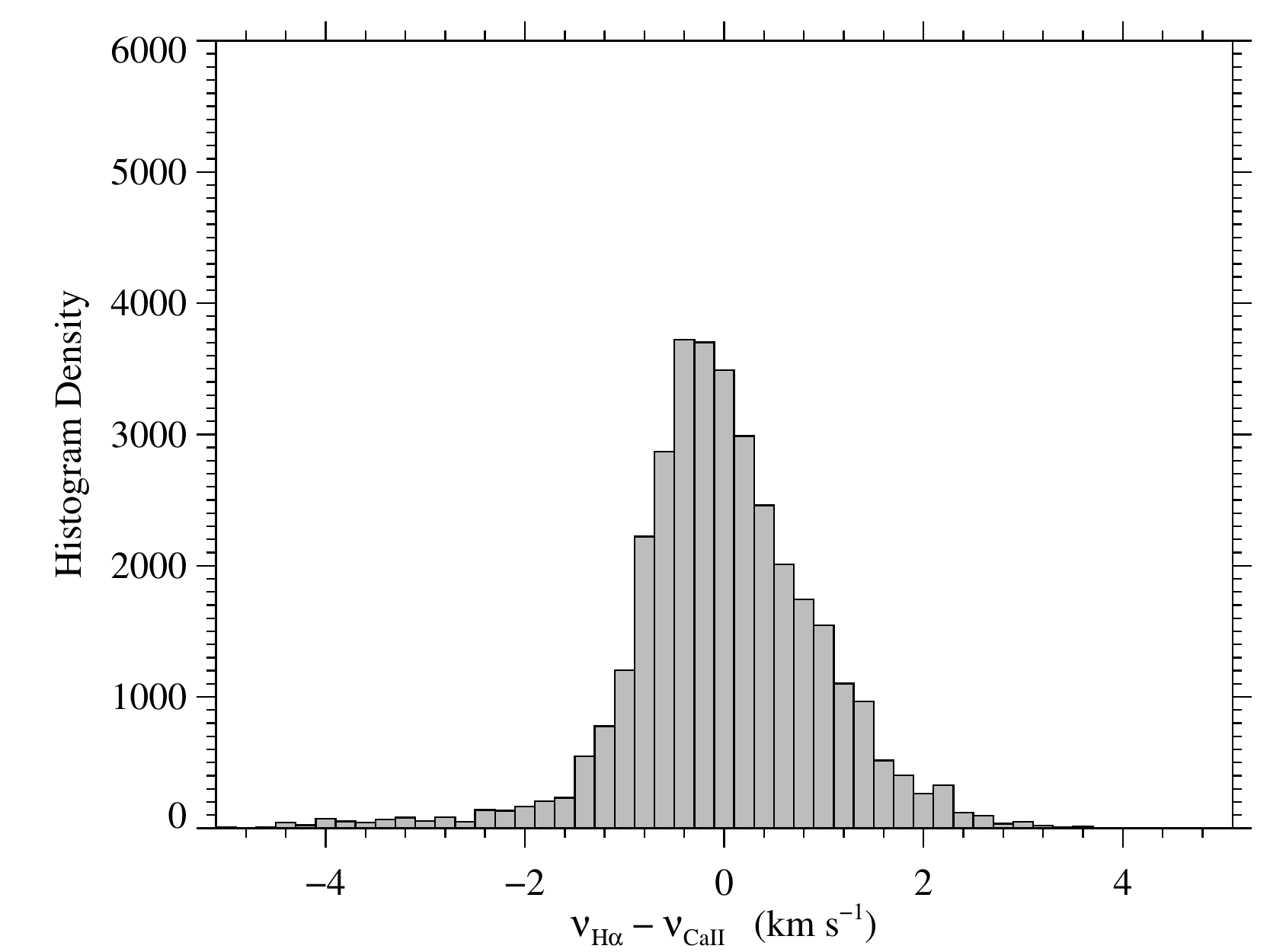}
    \end{subfigure}
    \begin{subfigure}{0.33\textwidth}
    \includegraphics[width=1\textwidth]{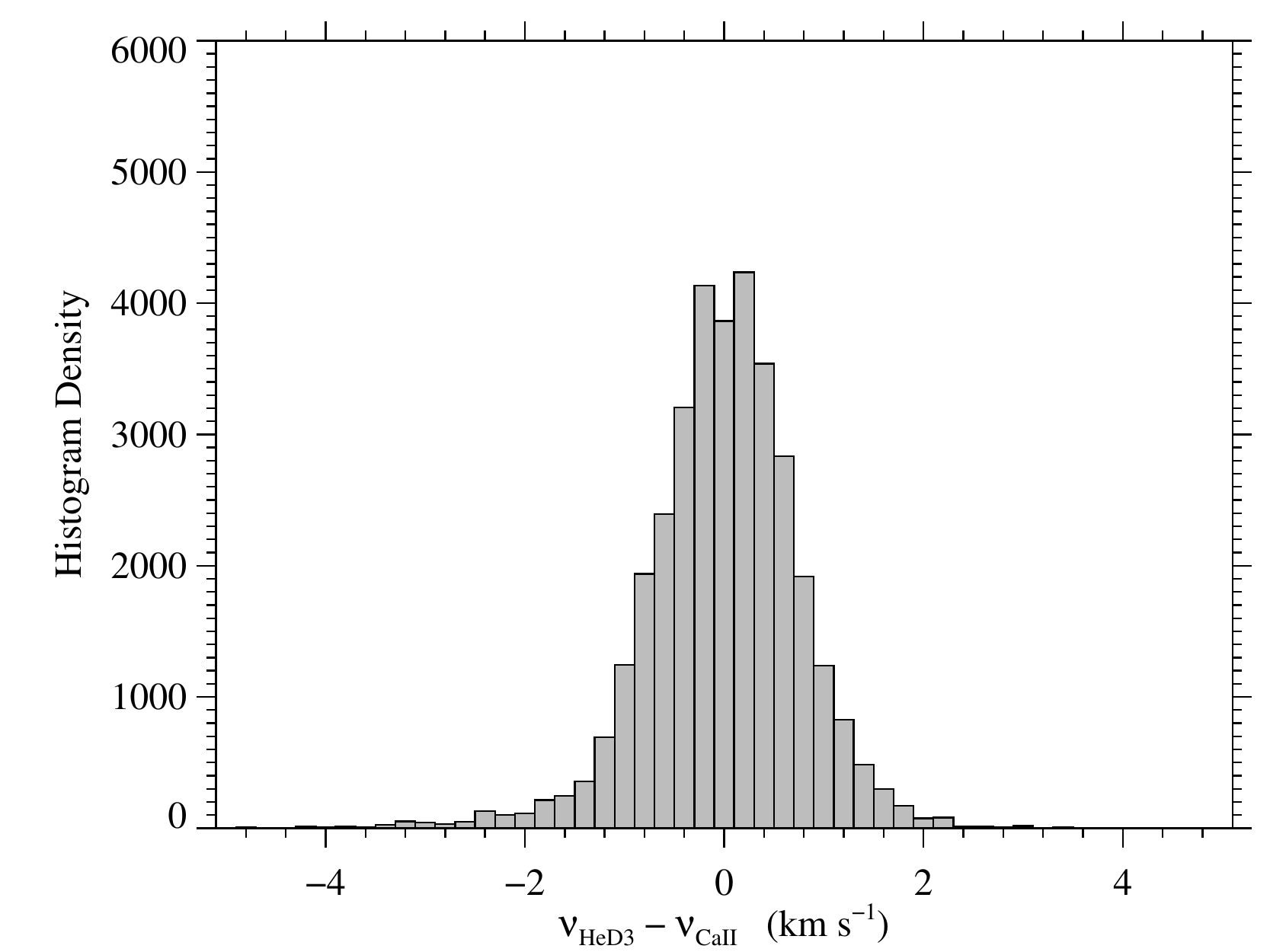}
    \end{subfigure}
    \begin{subfigure}{0.33\textwidth}
    \includegraphics[width=1\textwidth]{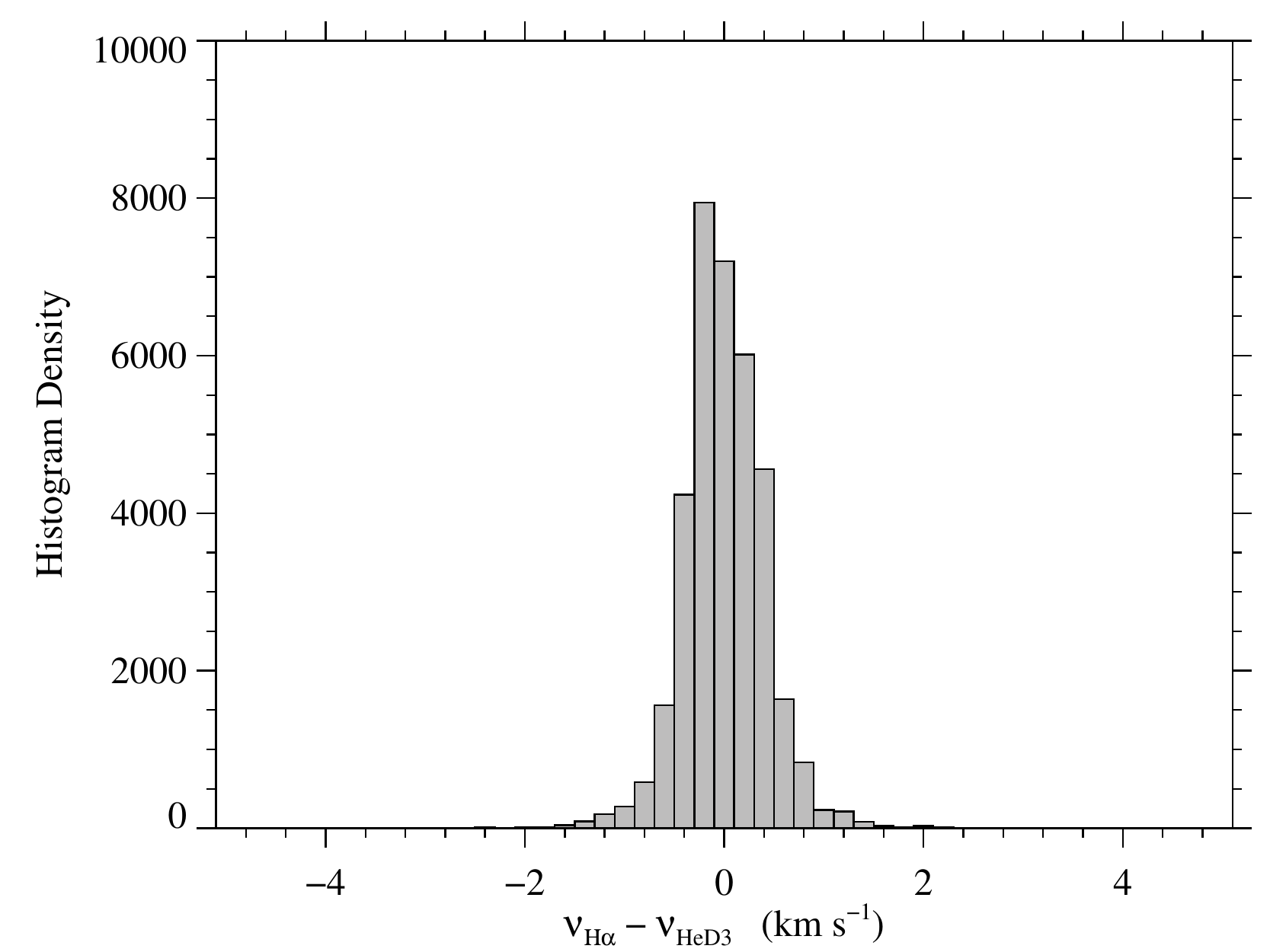}
    \end{subfigure}
    \caption{Histograms of the LOS velocity differences as measured with the observed spectral lines, $\nu_{H\alpha} - \nu_{CaII}$ (left), $\nu_{HeD3} - \nu_{CaII}$ (middle), and $\nu_{H\alpha} - \nu_{HeD3}$ (right).
    }
    \label{Fig:clusters_histo}
\end{figure*}

\subsection{Differential refraction}\label{Subsect:refraction}

Before comparing the velocities measured with the observed spectral
lines, it was necessary to evaluate the differential refraction produced by the
Earth's atmosphere. Differential refraction shifts the solar image in the zenithal direction
by a different amount depending on the wavelength, causing a different projection of the
solar image on the slit and leading to a different instantaneous spatial sampling of the 
prominence in the three observed wavelengths. For its calculation, we have used 
the time of the day, the average local temperature, pressure and humidity values during the
observing time interval and the position of the Sun relative to the horizon. At the time of the observations, the angle formed
by the slit and the zenithal direction was 38 degrees, which means that only a 60\% fraction of the
differential refraction takes place perpendicular to the slit, implying a relative displacement in
this direction of 0$\farcs$10 between \HeD3\ and \Halpha, and 0$\farcs$18 between \Halpha\ and \CaII. 
Since, both, the slit width and the scanning step in the 
direction perpendicular to the slit were 0$\farcs$75, the differential 
refraction shifts can be considered significantly smaller than the spatial resolution of our maps.  
Hence, we decided not to do any interpolation in the spatial (between scanning positions) nor temporal 
(between successive temporal spectral images at the same scanning position) directions. Finally, 
since the \CaII\ spectral data had a different spatial sampling along the slit, a linear 
interpolation of the \CaII\ velocity data was done in order to match the spatial sampling at this
wavelength with that of the other two datasets. The FOV along the slit was determined to pixel accuracy by comparing the 
intensity maps in the three spectral lines.

\section{Results}\label{Sect:results}

By comparing the three LOS velocity maps of scan position number 3 (upper panels in Fig.~\ref{Fig:Doppler}), it becomes evident that the plasma in the prominence exhibits an intense dynamical behaviour, with peak-to-peak variations larger than \hbox{5 km\,s$^{-1}$}. Similar results were obtained for all scan positions. There is a strong correlation between the neutral (\Halpha\ and \HeD3) and the ionised species (\CaII), with similar temporal and spatial variations in the three spectral ranges (blue/red velocities indicating plasma moving towards/away from the observer). This result is to be expected, since our selection criteria were addressed to get the three lines originated from the same volume within the prominence. It may be of interest to note that no clear evidence of short period oscillations (e.g, the typical three minute chromospheric period) has been detected. The most evident temporal variation corresponds to some fifty minutes.

\subsection{Velocity drifts between ionised and neutral plasma}\label{Sect:Drifts}

A detailed comparison of the Doppler velocities obtained for the ionised and the neutral species is crucial to detect their possible different dynamical behaviour. To that aim, we define the velocity drift as the deviation of the velocity measured in a neutral line, either \Halpha\ or \HeD3, and the velocity measured with the ionised line, the \CaII\ line. The neutral-ion drift velocities are expressed as \hbox{$w_1 = \nu_{H\alpha} - \nu_{CaII}$} and \hbox{$w_2 = \nu_{HeD3} - \nu_{CaII}$}. On the other hand, the difference in Doppler velocities between the neutral lines 
is defined as \hbox{$w_3 = \nu_{H\alpha} - \nu_{HeD3}$}.

The bottom panels in Fig.~\ref{Fig:Doppler} depict the drift velocities calculated for scan position 3 (same slit 
position as in the upper panels), $w_1$ (left panel), $w_2$ (middle panel), and $w_3$ (right panel). This figure already provides us with indications of the disparities in the dynamics between neutral and ionised plasma. The difference in Doppler velocities between the neutral lines ($w_3$) is close to zero or very small in certain regions, indicated by the white regions. Additionally, there are patches of blue and red indicating regions where the difference deviates significantly from zero. In contrast, the drift velocities between the neutral and ionised species ($w_1$ and $w_2$) are notably larger compared to the difference in velocities between the two neutral lines. Although there are still some regions where the difference is small, it is evident that a more pronounced contrast exists in the dynamics of the neutral and the ionised species of the plasma.

Fig.~\ref{Fig:clusters_histo} presents the histograms of $w_1$, $w_2$, and $w_3$ (left, middle, and right, respectively) for all the selected locations and all scan positions. The distribution of relative velocities is almost symmetric in all three histograms. However, the histogram corresponding to the velocity difference of the two neutral lines ($w_3$) is considerably narrower than those including the ions ($w_1$ and $w_2$), indicating that the dynamical behaviour of the two neutral species is more similar to each other and that the ion shows larger deviations. In numerous selected locations, \CaII\ exhibits large deviations compared to \HeD3\ and \Halpha. As a consequence, the larger widths of the histograms are the result of absolute values of ionized \CaII\ velocities larger than the neutral ones, similar to what has been found by other authors  \citep[e.g.,][]{Khomenko2016, Wiehr2019, Wiehr2021}. Most of the drift excess of \CaII\ is below 1\,km\,s$^{-1}$.

\begin{figure}[!t]
\sidecaption
  \includegraphics[width=\hsize]{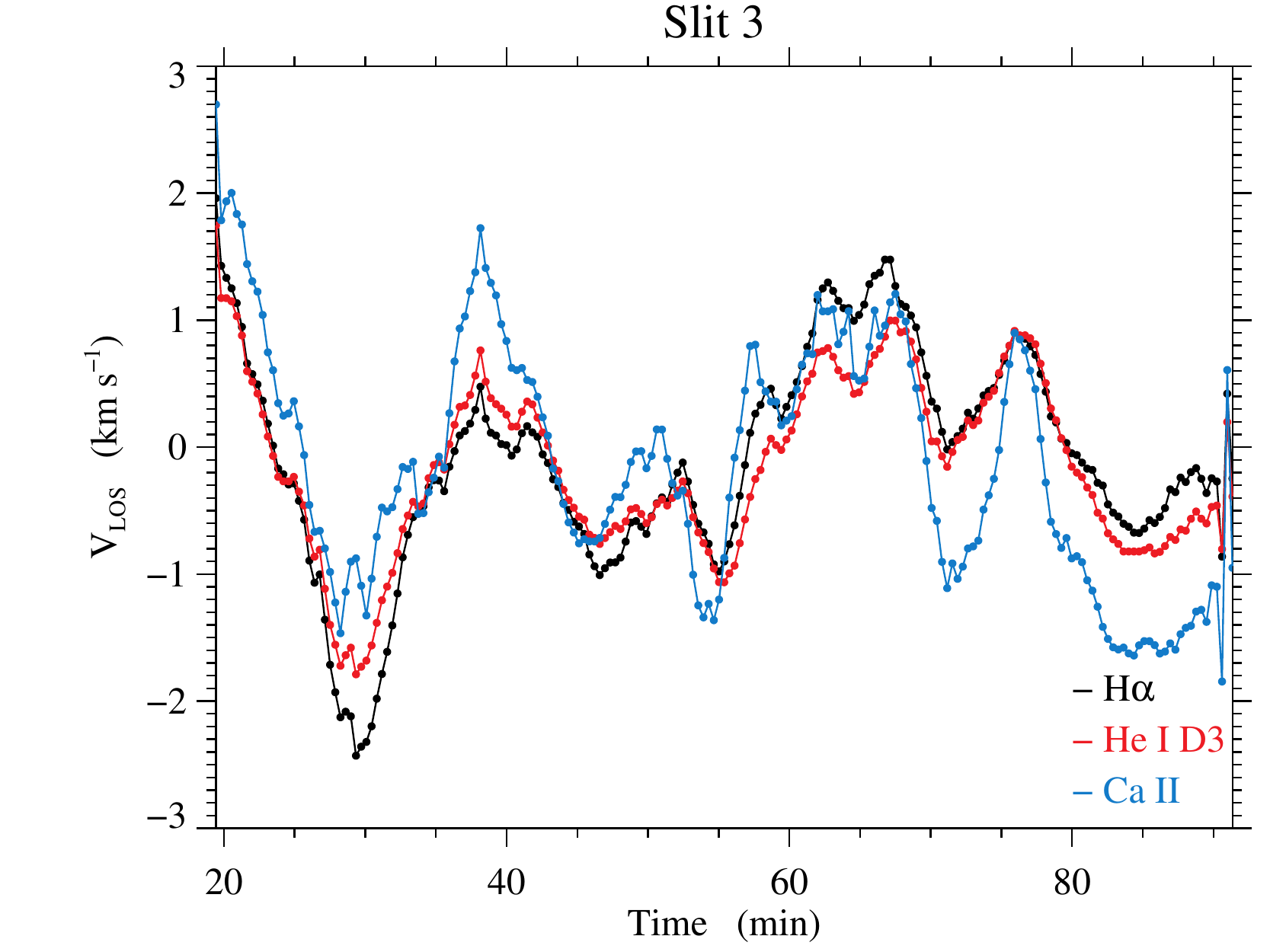}
  \caption{LOS velocity measured with the observed spectral lines at the fixed position shown as a black line in the bottom plots of Fig.\,\ref{Fig:Doppler}.
  } 
  \label{Fig:v_variation}
\end{figure}

To further investigate the temporal variation of the LOS velocities, we present a specific slit position marked by the vertical black line in Fig. \ref{Fig:Doppler}. At this fixed slit position, the temporal evolution of the LOS velocities for the three different spectral lines is shown in Fig. \ref{Fig:v_variation}. The LOS velocities of \CaII\ generally follow the overall trend exhibited by the neutral species with distinct moments of deviation and subsequent recovery. For instance, a notable recovery in the \CaII\ LOS velocities around minutes 20, 30, and 55. Conversely, deviations from the general trend are evident at minutes 40, 70, and 85. This is in line with the results of previous works \citep[e.g.,][]{Khomenko2016}, with ions showing larger velocity excursions than neutrals, accompanied with moments with small velocity drifts. Slight differences between the velocities of \HeD3\ and \Halpha are also observed in particular moments (minutes 30 and 80) indicating that the coupling between the neutral species is also not complete.

\begin{figure*}
    \centering
    \begin{subfigure}{0.33\textwidth}
    \includegraphics[width=1\textwidth]{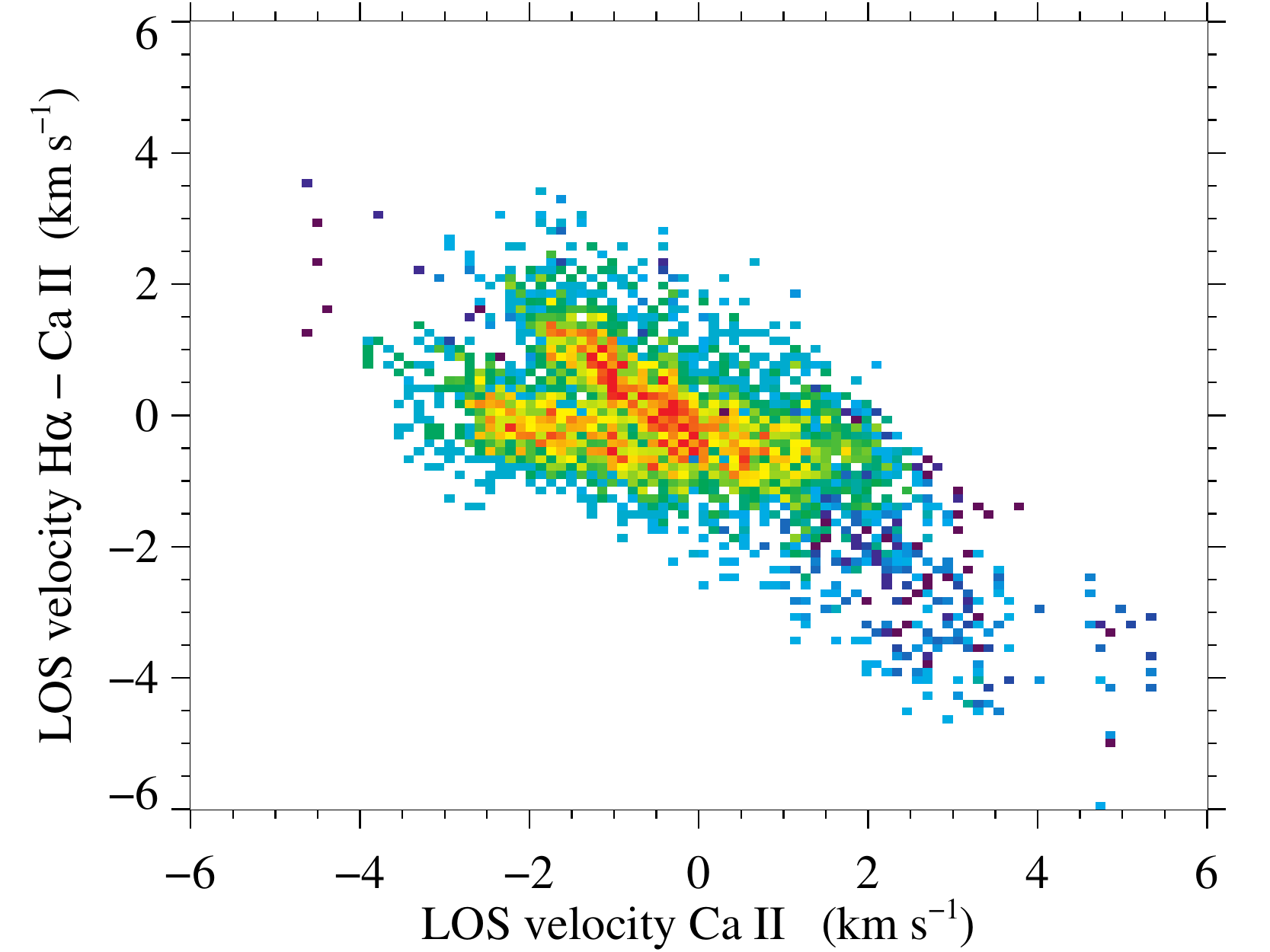}
    \end{subfigure}
    \begin{subfigure}{0.33\textwidth}
    \includegraphics[width=1\textwidth]{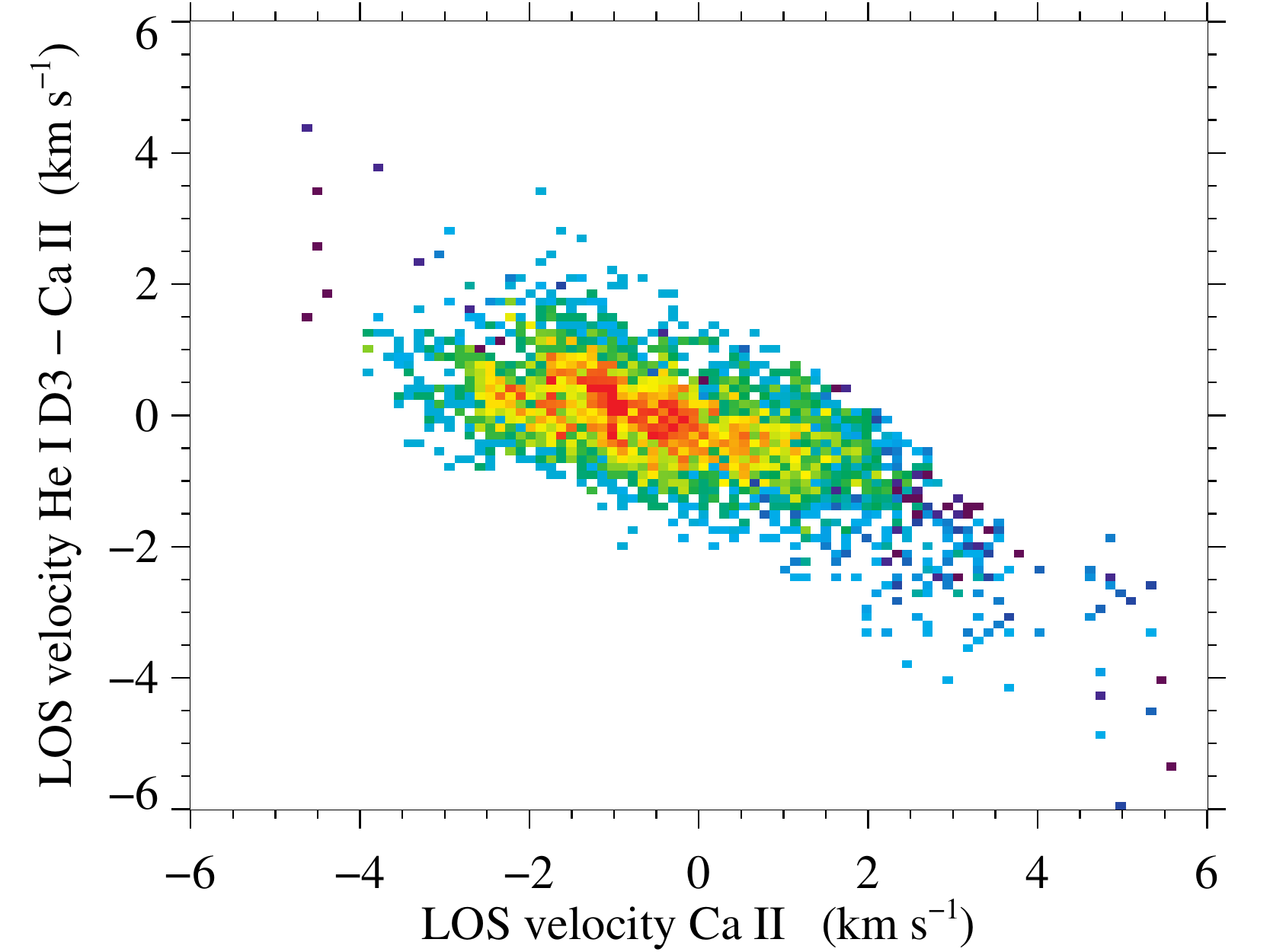}
    \end{subfigure}
    \begin{subfigure}{0.33\textwidth}
    \includegraphics[width=1\textwidth]{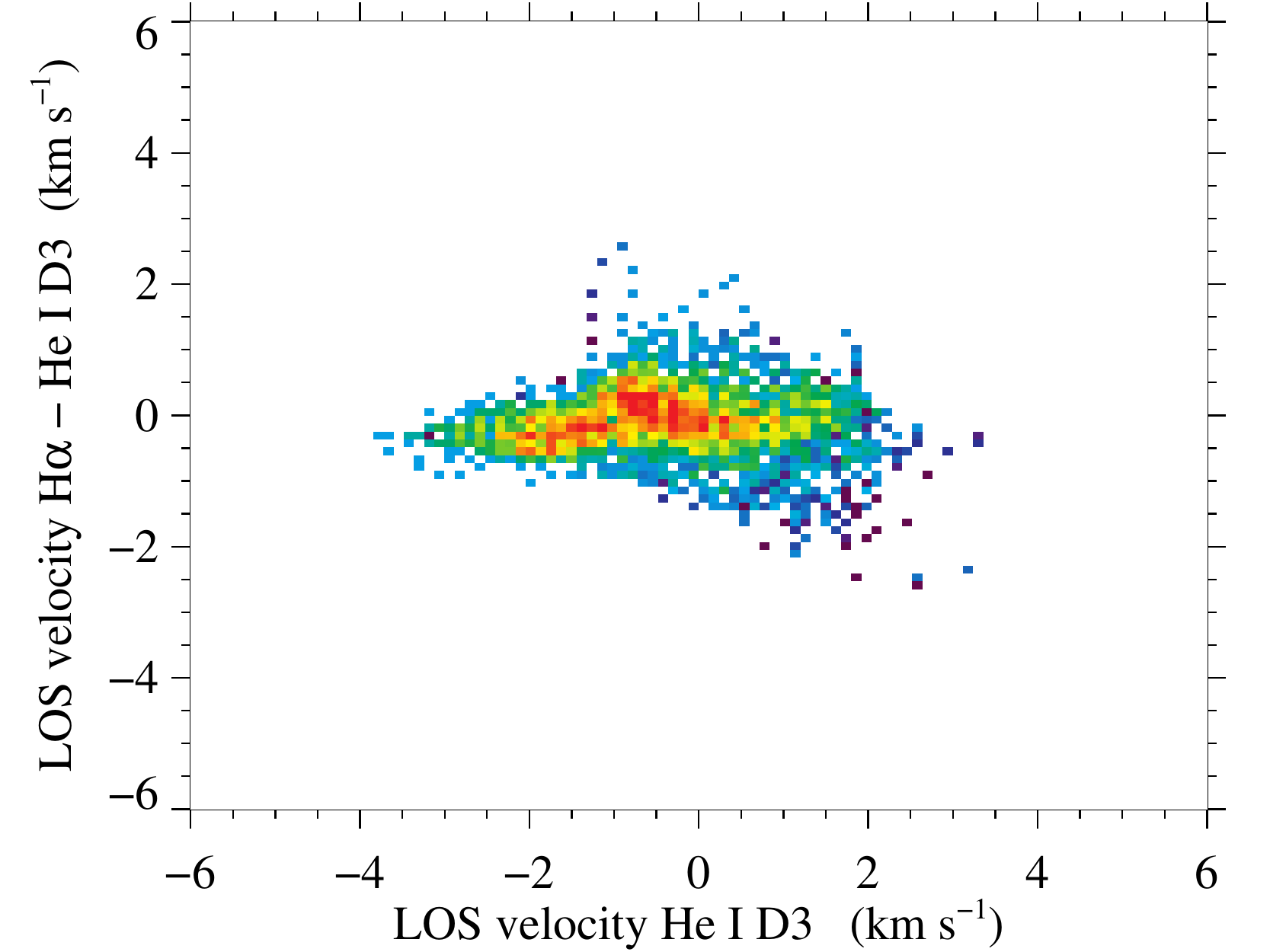}
    \end{subfigure}
    \begin{subfigure}{0.33\textwidth}
    \includegraphics[width=1\textwidth]{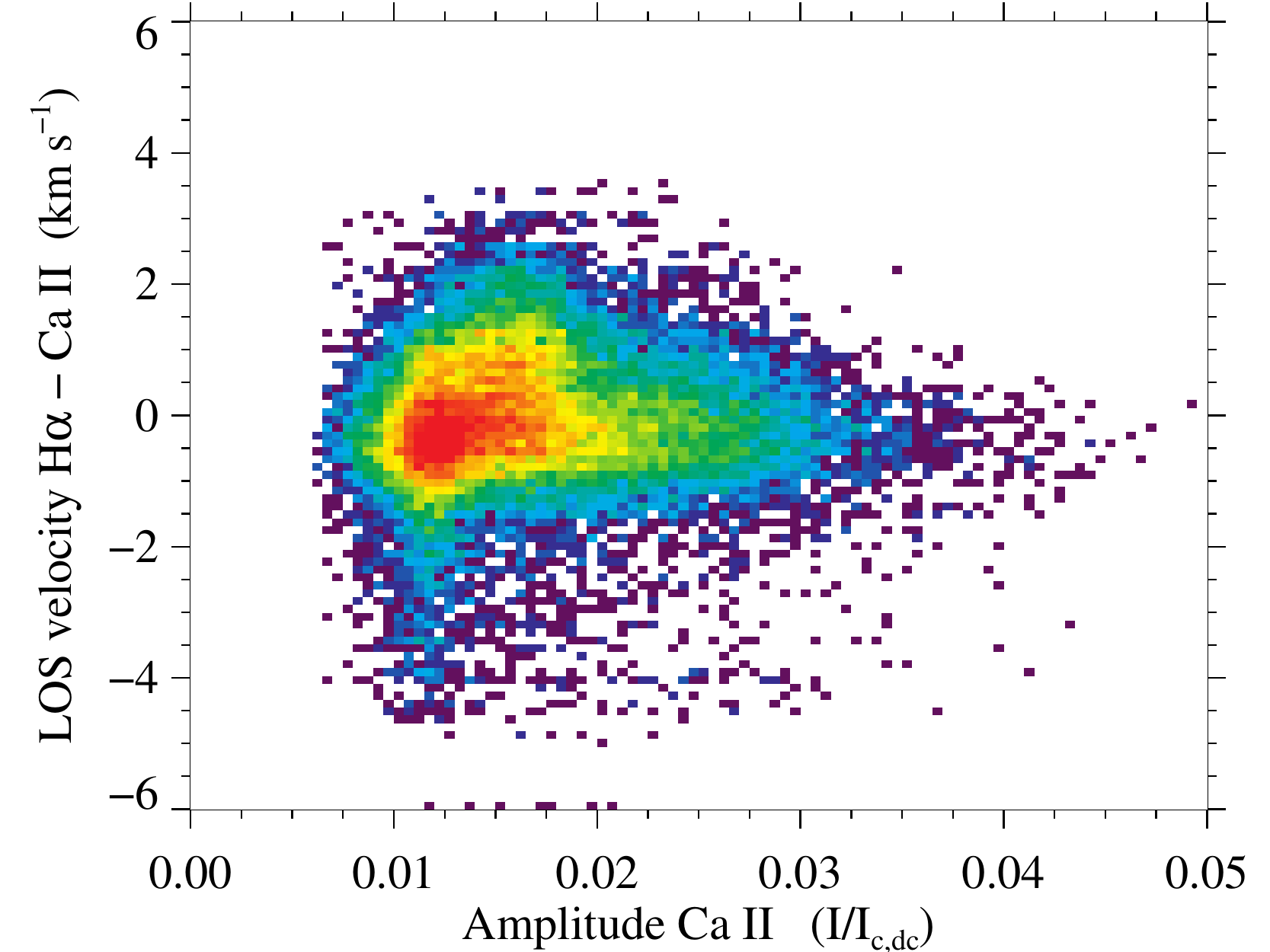}
    \end{subfigure}
    \begin{subfigure}{0.33\textwidth}
    \includegraphics[width=1\textwidth]{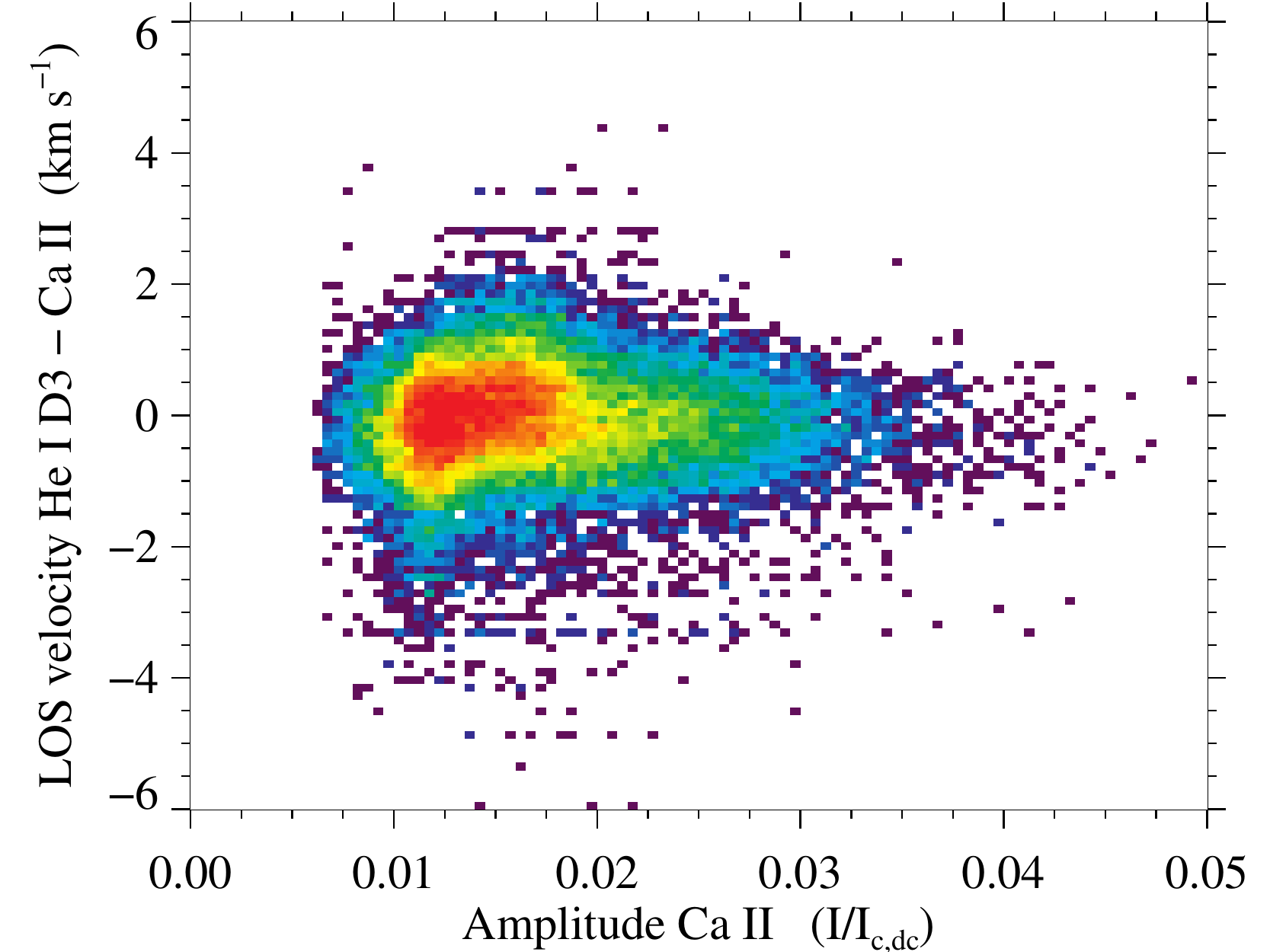}
    \end{subfigure}
    \begin{subfigure}{0.33\textwidth}
    \includegraphics[width=1\textwidth]{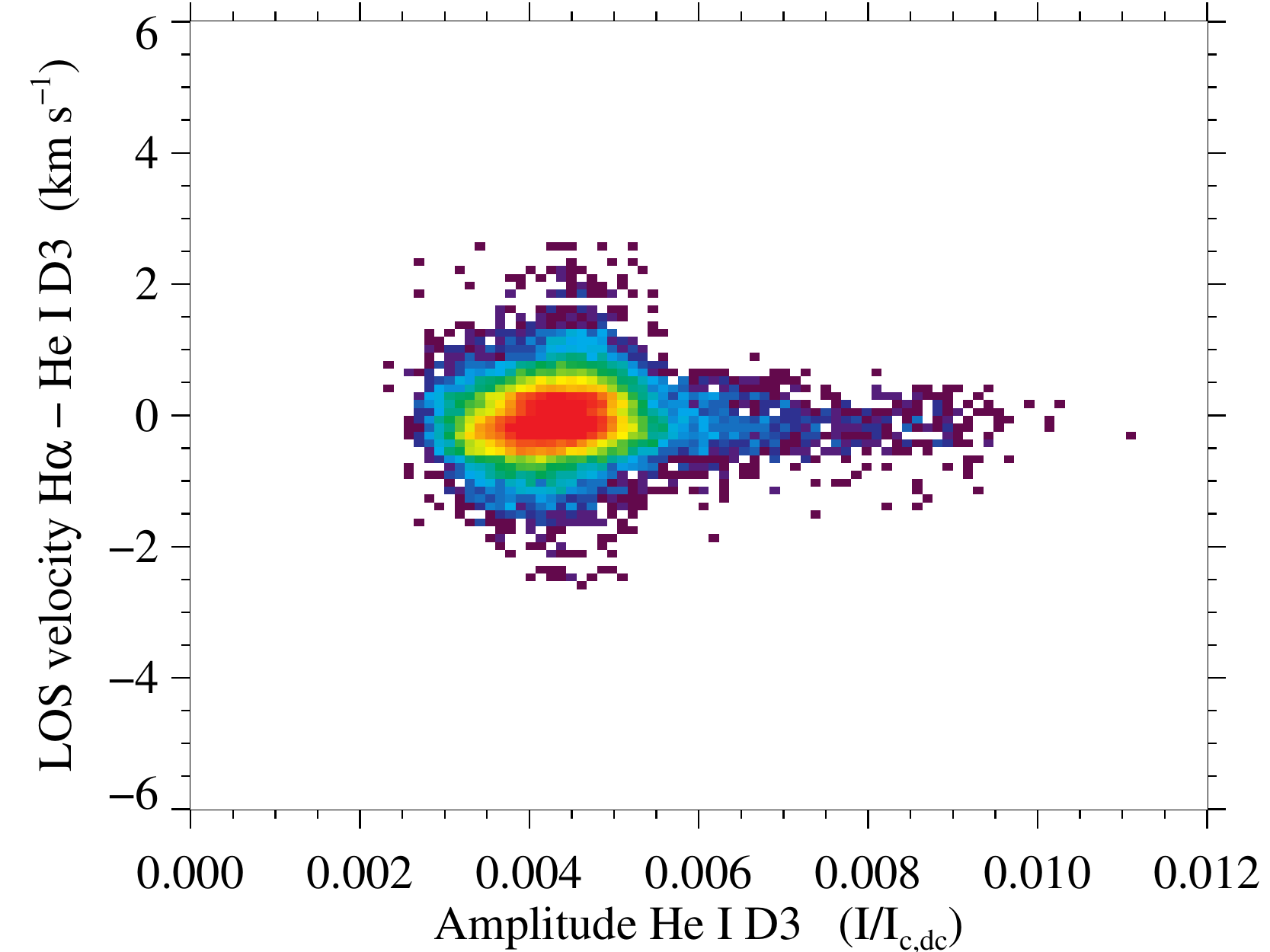}
    \end{subfigure}
    \caption{\emph{Top}: Two-dimensional histograms showing the dependence of the velocity drifts with respect to a reference velocity. \emph{Bottom}: Two-dimensional histograms showing the dependence of the velocity drifts with respect to a reference spectral line amplitude. See Sect.\,\ref{Sect:Drifts} for the explanation.}
    \label{Fig:clusters_}
\end{figure*}

A more detailed information about the drift velocity distributions shown in Fig\,\ref{Fig:clusters_histo} can be obtained by looking at the dependence of each particular drift bin of the histograms with the LOS velocity itself or the amplitude of the profiles. Since all LOS velocities are rather similar according to the top panels of Fig.\,\ref{Fig:Doppler}, the particular LOS velocity used as a reference is not relevant. For the neutral-ion drifts, we have chosen the \CaII\ velocity, and the \HeD3\ velocity for the neutral-neutral drift. The resulting bi-dimensional histograms (number of points with a given drift velocity bin and a given reference LOS velocity bin) are presented in the top panels of Fig.\,\ref{Fig:clusters_}. A thermal-like lookup table has been used with red/blue indicating the largest/smallest clustering of points. Despite the scatter of the histogram points, a linear relationship can be seen for the two neutral-ion plots (left and middle panels). Negative/positive LOS velocities are accompanied by positive/negative drifts $w_1$ and $w_2$, implying that the ion absolute velocities are larger than that of the neutrals (i.e., when there is a blue/red shift, the \CaII\ spectral line is more blue/red shifted). This conclusion is similar to that reached by \citet{Wiehr2019, Wiehr2021} using different sets of spectral lines of neutral and ionised species. In contrast, the neutral-neutral 2D histogram (rightmost top panel) does not show that tendency: the neutral-neutral drift velocity is independent of the total velocity.

The bottom panels of Fig.~\ref{Fig:clusters_} present the bi-dimensional histograms of the drift velocity, now using the observed amplitude as the reference parameter. As for the case of velocities, it is not relevant the amplitude of which line is used in particular, since our selection criteria ensure a linear relationship between the three amplitudes to avoid opacity effects in the line formation (see Fig.\,\ref{Fig:amplitudes}). As for the velocities, we have used \CaII\ amplitudes as a reference for the analysis of the dependence of the neutral-ion drifts and \HeD3\ amplitudes for the neutral-neutral drifts. Now, the 2D histograms do not show any particular trend, as expected. The bins are equally distributed for the positive and negative drifts. It is to be remarked, though, that the neutral-ion drift velocities (left and middle panels) show a larger scatter for smaller amplitudes. This behaviour can be understood in terms that a smaller particle density (lower amplitude) also encompasses a smaller collisional frequency and, consequently, a smaller coupling. This result reinforces our criterion to discard those line profiles with clear indications of line saturation and keep only those points in the FOV in the optically thin regime.

\subsection{Unresolved velocities}

In addition to the analysis of the spatially resolved LOS velocities described in the previous sections, there also exist small-scale motions whose main effect is to broaden the observed spectral lines. These motions may have a thermal nature, in which case they are usually parametrised by their thermal Doppler velocity \hbox{$\Delta v_{\rm th} = \sqrt{2k_B T_{\rm kin}/m}$}, where $k_B$ is the Boltzmann constant and $m$ the mass of the considered atomic species, and whose value determines the kinetic temperature, $T_{\rm kin}$. Unresolved motions may also exist with a non-thermal origin and can be caused by small-scale velocity gradients in the LOS direction, turbulent motions, high-frequency waves, etc. All these effects are often taken into account via a non-thermal velocity, $\Delta v_{\rm nth}$, that is usually added quadratically to the thermal motions, so that the total Doppler velocity, $\Delta v_{\rm D}$, is

\begin{align} \label{eq:Doppler}
\Delta v_{\rm D} = \sqrt {(\Delta v_{\rm th})^2+ (\Delta v_{\rm nth})^2},
\end{align}

\noindent
leading to the broadening of a spectral line in terms of wavelength, $\Delta \lambda_D$, as given by the equation

\begin{align} \label{eq:linewidth}
    \Delta \lambda_D = \frac{\lambda_0}{c}\sqrt{(\Delta v_{\rm th})^2+ (\Delta v_{\rm nth})^2} = 
        \frac{\lambda_0}{c} \sqrt{\frac{2k_B T_{\rm kin}}{m} + (\Delta v_{\rm nth})^2},
\end{align}

\noindent
where $\lambda_0$ is the central wavelength of the observed spectral line and $c$ is the speed of light.

It is interesting to note that the thermal velocity component in Eq.\,\ref{eq:linewidth} depends on the mass $m$, while the non-thermal velocity $\Delta v_{\rm nth}$ does not. By measuring the width of spectral lines of different elements (i.e., different atomic mass), it is feasible to separate the two components. A number of authors have used this approach to derive the kinetic temperature, $T_{\rm kin}$, in prominences and the amplitude of the non-thermal velocity component $\Delta v_{\rm nth}$ \citep[][and references therein]{Ramelli_etal_2012, Park_etal_2013, Wiehr_etal_2013, Stellmacher_Wiehr_2015, Stellmacher_Wiehr_2017, Okada_etal_2020}. Since our dataset is formed by spectral lines coming from three different atomic species (hydrogen, helium and calcium), we have applied the same analysis to our observed prominence. 

For the interpretation of the line width, the optical thickness of the prominence region where the spectral line is formed also plays a critical role. Optical thickness affects mainly the core of the line, where self-absorption along the line-of-sight is larger, and leads to an emission line core flatter than would be obtained in an optically thin scenario. Even a self-reversal core may appear if the optical thickness is very large \citep[see, v.g.,][]{Park_etal_2013}. As a result, spectral lines formed in an optically non-thin environment are wider than expected. To account for this additional broadening effect, some authors have used the model of optically thick slab \citep[see, v.g.,][]{Park_etal_2013, Okada_etal_2020, Jejcic}. This simple model makes it possible to determine $T_{\rm kin}$ and $\Delta v_{\rm nth}$ under the assumption that both parameters are constant along the line of sight and that there are no bulk velocity gradients. Typical values for these two parameters range from 4000 to \hbox{20000 K} for $T_{\rm kin}$ and from 3 to \hbox{20 \kms} for $\Delta v_{\rm nth}$ \citep[see][and references therein]{Okada_etal_2020}. 

In general, the opacity will not be the same for all the observed spectral lines and the plasma volume sampled by each line will be different. In this case, and if the homogeneity condition can not be applied, volumes with different physical conditions will lead to a different line broadening and the results obtained from a number of spectral lines may not be directly comparable. For instance, \citet{Okada_etal_2020} conclude that the \CaIIK\ line is very much affected by the optical thickness and is not suitable for this type of analysis. It is not surprising, thus, that the values retrieved for the pair $T_{\rm kin}$ and $\Delta v_{\rm nth}$ depend on the particular choice of spectral lines used for their evaluation if a deep analysis of the consequences of opacity is not performed. In our case, we have restricted our study to those points in the prominence that have been considered as optically thin and without significant velocity gradients, according to the criteria described in Sect.\,\ref{Subsect:k-means}. This way, we can expect that the observed spectral lines are formed over the whole line of sight and all the LOS variations of the physical parameters are smoothed. We cannot exclude, however, that the formation of certain spectral lines may have a different weight in regions of larger/lower temperature/density, thus biasing the calculation towards the properties of those regions. Nevertheless, with all the selection criteria that we have applied, we expect to have minimised the impact of all the inhomogeneities along the line of sight. 

\begin{figure}[!t]
\sidecaption
  \includegraphics[width=\hsize]{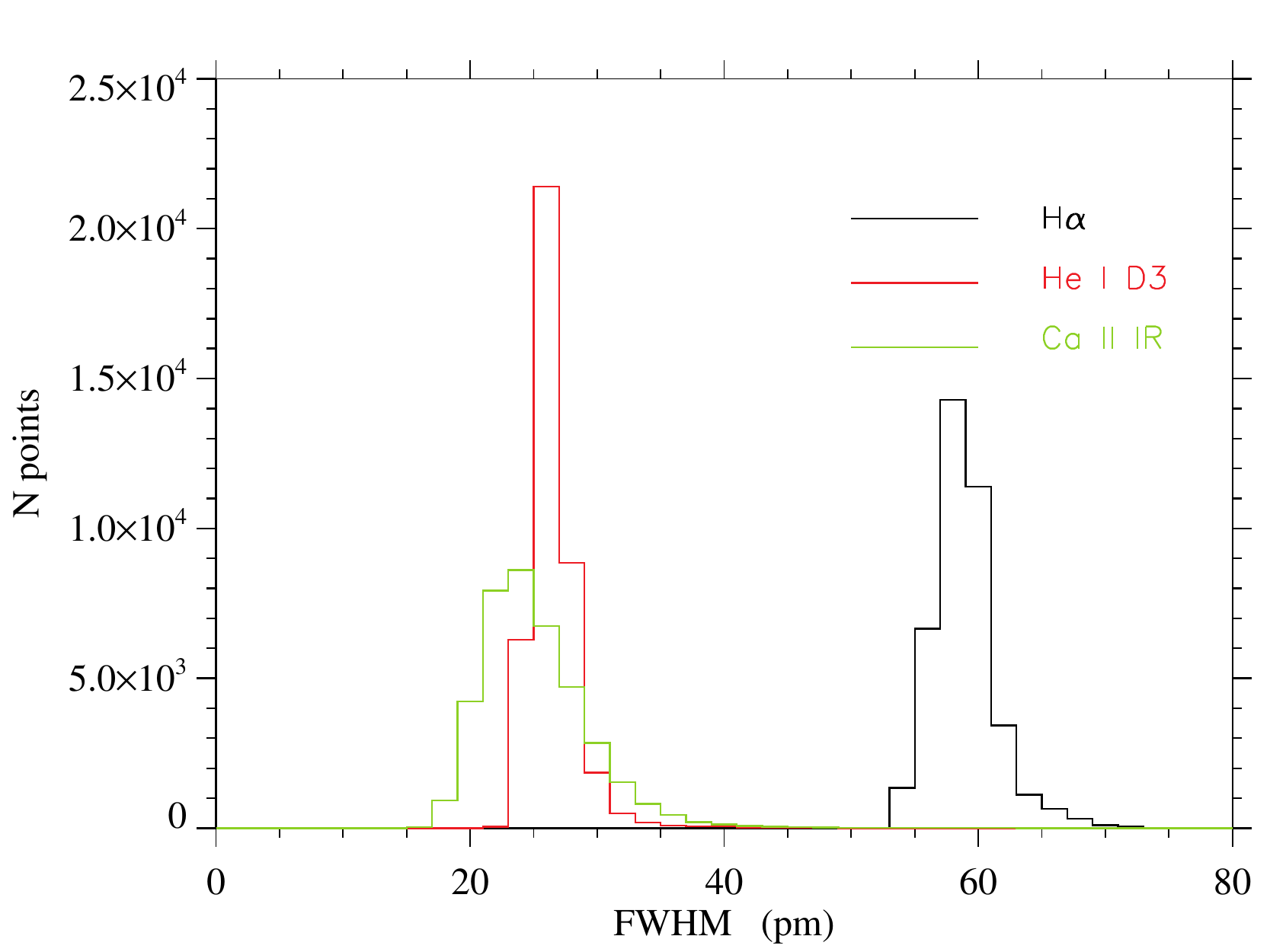}
  \caption{Histograms of the full-width at half-maximum (FWHM) of the emission profiles at the selected points. The average values and standard deviations of the distributions for each spectral line are presented in Table \ref{tab:FWHM_dld}. The inserted legend indicated the color assigned to each spectral line.}
  \label{Fig:line_widths}
\end{figure}

\begin{table}
    \centering
    \begin{tabular}{l|c|c|c}
    \hline
    Parameter                                 & \Halpha          & \HeD3           & \CaII \\
    \hline
FWHM (pm)                                     & $60 \pm 3$       & $28 \pm 2$      &  $26 \pm 4$ \\
$\Delta\lambda_D$ (pm)                        & $35 \pm 2$       & $16 \pm 1$      &  $16 \pm 3$ \\
$\Delta v_D$ (\kms)                           & $15.9 \pm 0.7$   & $8.1 \pm 0.5$   &  $5.5 \pm 0.9$ \\
$T_{\rm kin}\ (\Delta v_{\rm nth}= 0)$ (kK)   & $15.3 \pm 1.4$   & $15.8 \pm 1.8$  &  $74 \pm 24$ \\
    \hline
    \end{tabular}
    \caption{Average values of the full-width at half-maximum (FWHM), total Doppler width ($\Delta\lambda_D$), the corresponding total Doppler velocity of unresolved motions ($\Delta v_D$) and the kinetic temperature ($T_{\rm kin}$) that is derived in case the non-thermal unresolved velocity is zero.}
    \label{tab:FWHM_dld}
\end{table}

\begin{figure*}
    \centering
    \begin{subfigure}{0.33\textwidth}
    \includegraphics[width=1\textwidth]{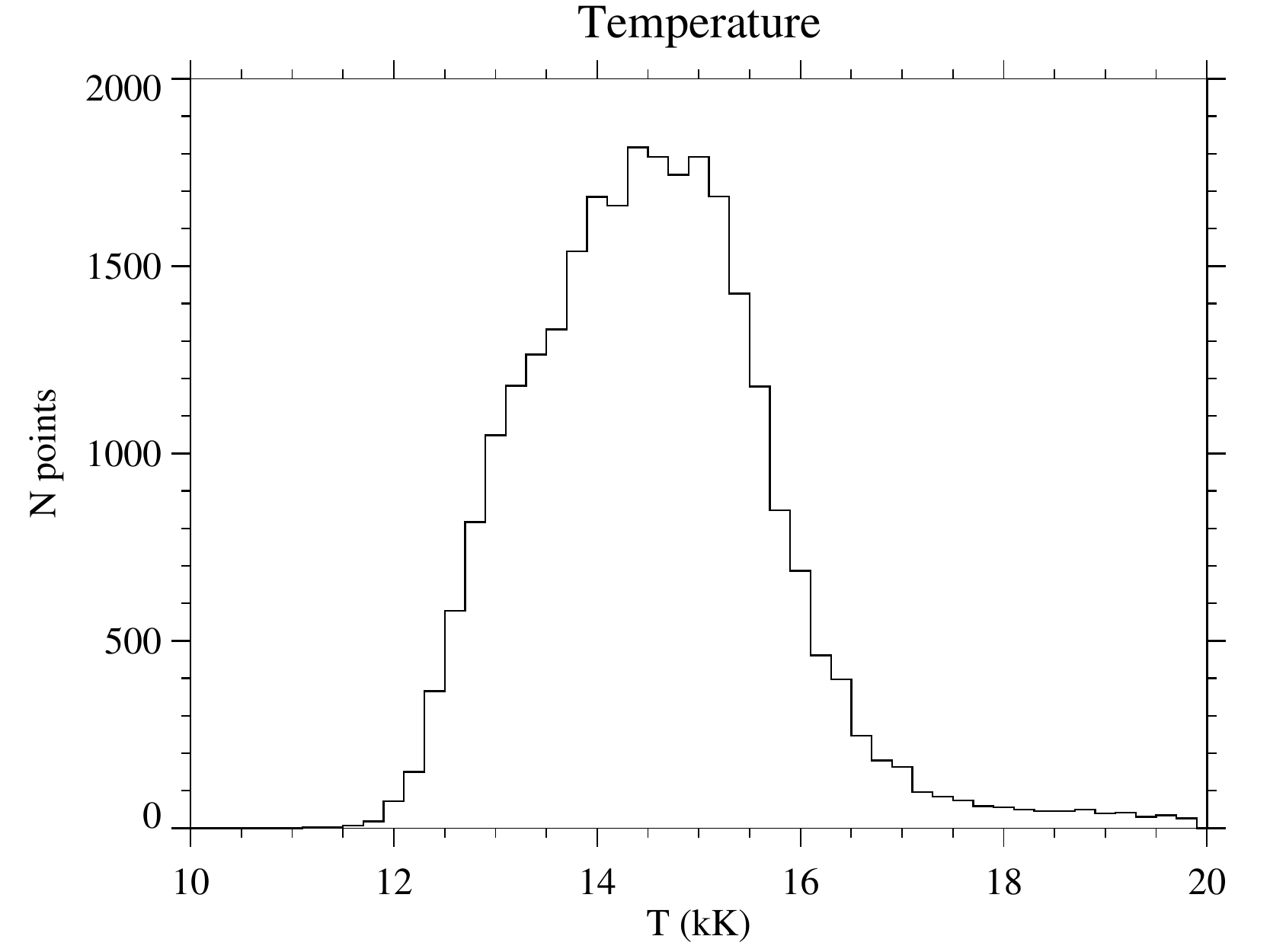}
    \end{subfigure}
    \begin{subfigure}{0.33\textwidth}
    \includegraphics[width=1\textwidth]{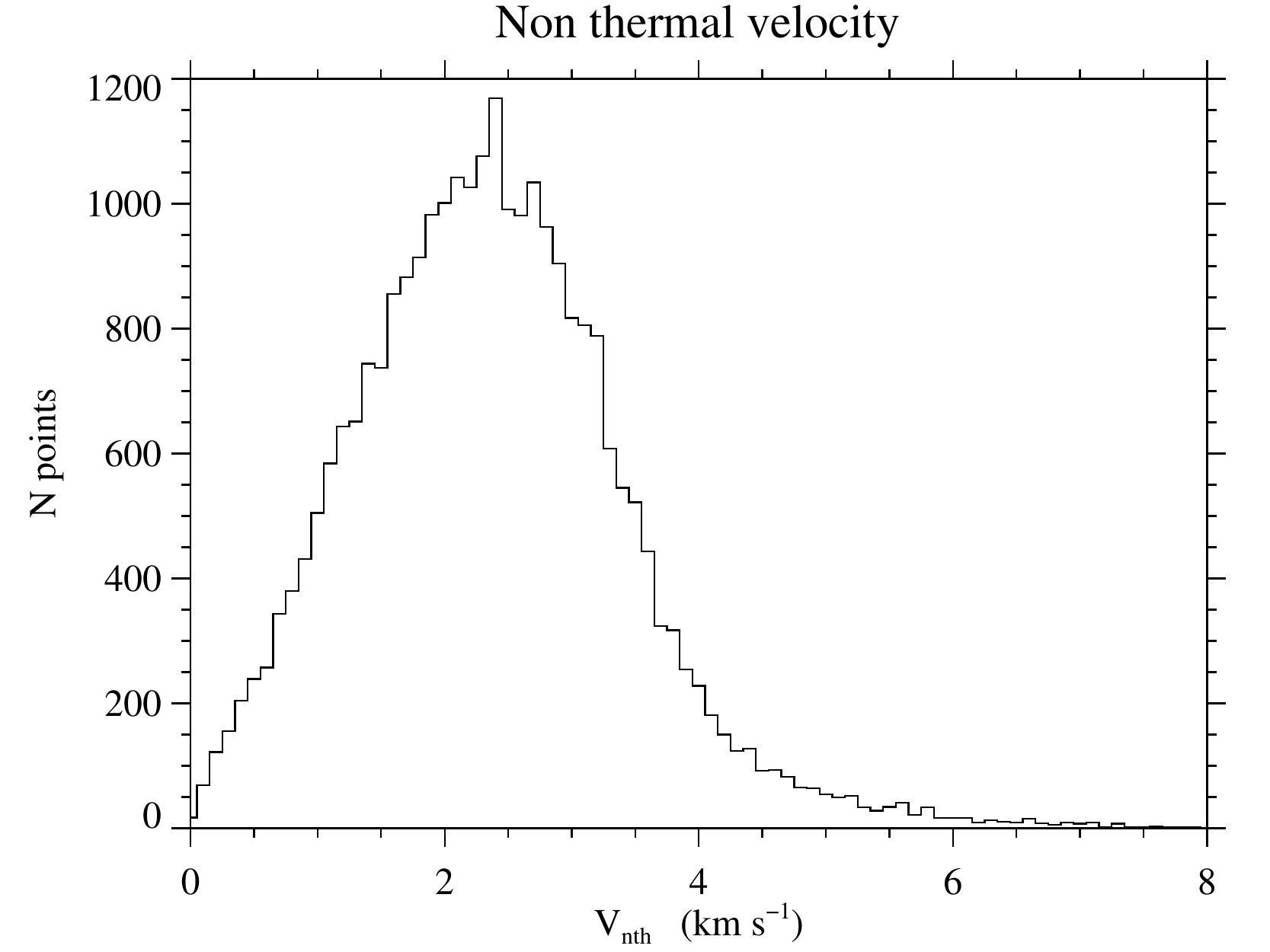}
    \end{subfigure}
    \begin{subfigure}{0.33\textwidth}
    \includegraphics[width=1\textwidth]{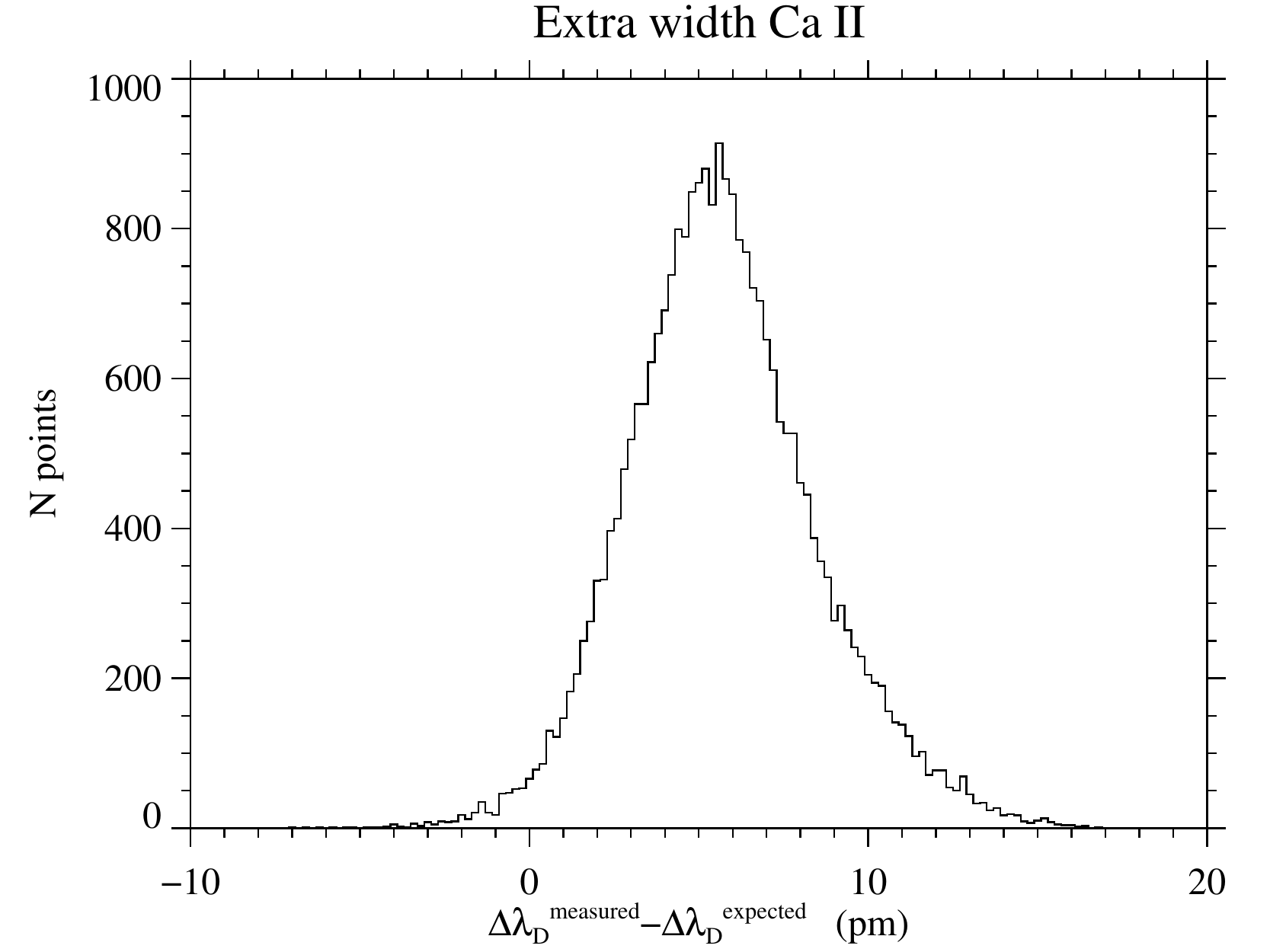}
    \end{subfigure}
    \caption{Distribution of kinetic temperatures (\emph{Left}) and non-thermal Doppler width (\emph{Centre}) derived from the \Halpha\ and \HeD3\ spectral lines in all selected points using Eq.\,\ref{eq:linewidth}. With these parameters, the expected Doppler width has been calculated for the \CaII\ spectral line at every point and the excess (measured minus expected) non-thermal velocities histogram is presented (\emph{Right}).
    }
    \label{Fig:Ca_deviation}
\end{figure*}
For the calculation of $\Delta \lambda_D$, we have first measured the full-width at half-maximum (FWHM) of the emission profiles at the selected points. The histogram of the obtained values are represented in Fig.\,\ref{Fig:line_widths} and the average values and standard deviations of the distributions for each spectral line are presented in Table \ref{tab:FWHM_dld}. It has to be noted that the histograms are slightly asymmetric and the average values are slightly larger than the peak values (probably due to some residual profiles that are affected by the opacity). Following the same argument as for the velocities, we can estimate the FWHM errors to be smaller than a pixel (even though error propagation gives much smaller values), which correspond, approximately, to \hbox{$\pm\,0.18$ pm} in \Halpha, \hbox{$\pm\,0.16$ pm} in \HeD3, and \hbox{$\pm\,0.32$ pm} in \CaII. These uncertainties are one order of magnitude smaller than the standard deviation of the distributions and, probably, indicate that the width of the histograms is given by real changes in the physical properties of the different observed points. Such variations are expected since our selected field-of-view is close to the prominence-corona interface and it is well known that a number of instabilities may appear there \citep{Khomenko2014, Hillier2018, Popescu2021a, Popescu2021b, MartinezGomez2022, MartinezSykora2023}.

For the conversion from FWHM to Doppler width, we have taken into account the fine structure of the \Halpha\ and \HeD3\ spectral lines. To that aim, we have assumed that each fine-structure component contributes to the observed spectral line with a Gaussian emission profile with the wavelengths and relative amplitudes listed in the NIST database \citep{NIST_table}. This calculation is compatible with the optically thin, symmetric and single-peaked profiles that we have selected from our dataset. Finally, we have computed the FWHM of the resulting spectral line for different values of $\Delta\lambda_D$ calculated from a set of kinetic temperatures in the range from 5000 to 25000 K and zero non-thermal velocity. A polynomial fit gives the following results for \Halpha\ and \HeD3: 

\begin{align} \label{eq:FWHM2dld_Halpha}
    \Delta\lambda_{D,{\rm H}\alpha} = 0.627 \cdot {\rm FWHM_{\rm H\alpha}} -2.76
\end{align}

\noindent
and

\begin{align} \label{eq:FWHM2dld_He}
    \Delta\lambda_{D,{\rm He}} =-0.00564 \cdot {\rm FWHM}_{\rm He}^2 + 0.805 \cdot {\rm FWHM_{He}} -1.97,
\end{align}

\noindent
where both parameters, $\Delta\lambda_{D}$ and FWHM, are expressed in pm units in both equations. Eqs. \ref{eq:FWHM2dld_Halpha} and \ref{eq:FWHM2dld_He} have an accuracy in $\Delta\lambda_{D}$ of \hbox{$\pm$ 0.07 pm} (\Halpha) and \hbox{$\pm$ 0.02 pm} (\HeD3). It has to be noted that both equations give an unphysical negative value for $\Delta\lambda_{D}$ for zero FWHM (i.e., zero temperature). So, care has to be taken for not using them outside the interval used to calculate the fit. 

For \CaII, the standard conversion for a single gaussian has been used:

\begin{align} \label{eq:FWHM2dld_Ca}
    \Delta\lambda_{D,{\rm Ca}} = \frac{{\rm FWHM}_{\rm Ca}}{2\sqrt{\ln{2}}} = \frac{{\rm FWHM}_{\rm Ca}}{1.665},
\end{align}

As a first attempt to see the compatibility of the three Doppler widths, we have computed the corresponding kinetic temperatures assuming that the line broadening has exclusively a thermal origin $(\Delta v_{\rm nth}= 0)$. The result also appears listed in Table \ref{tab:FWHM_dld}. It is remarkable how close the kinetic temperatures derived for the two neutral species (hydrogen and helium) are. This means their Doppler widths are mainly determined by thermal motions, since their values are closely scaled with the inverse of the square root of their masses.  However, the ionised species (calcium) has almost the same Doppler width as helium and requires a much larger kinetic temperature. So, calcium atoms have either a larger kinetic temperature or a larger non-thermal contribution. In whatever case, the conclusion is that the ionised species has larger non-resolved velocities than the neutral ones. In a scenario in which all non-resolved motions have a thermal origin, and lighter/heavier species move faster/slower, calcium atoms would require a kinetic temperature five times larger than that of hydrogen or helium ones. A mechanism to rise the kinetic temperature of the ions to such a large value would then be required. 

To further evaluate the potential differences of the ion behaviour, we have calculated the $T_{\rm kin}$ and $\Delta v_{\rm nth}$ values applying Eq.\,\ref{eq:Doppler} to \Halpha\ and \HeD3\ at every selected point. The left and central histograms of Fig.\,\ref{Fig:Ca_deviation} display the obtained distributions. The distribution peaks around \hbox{14.5 kK} for $T_{\rm kin}$ (slightly below the numbers show in Table \ref{tab:FWHM_dld}, since, as expected, part of the unresolved motions of the neutrals are now considered to have a non-thermal origin by the fit) and \hbox{2.5 \kms} for $\Delta v_{\rm nth}$. With the retrieved parameters, we have calculated the expected Doppler width for the \CaII\ spectral line at every point. The excess (measured minus expected) non-thermal velocities histogram is presented on the right of Fig\,\ref{Fig:Ca_deviation}. We can see that an excess of 5-6 pm is needed on average to explain the measured Doppler width of \CaII, with a distribution that is clearly biased towards positive values. We cannot say whether this Doppler width excess has a thermal or a non-thermal origin (or a combination of both). But, in any case, as with the bulk velocities, the results show that ions tend to move faster also at non-resolved scales.

\section{Discussion and Conclusions} \label{Sect:discus_conclus}

In this work, we have presented evidences that ions and neutrals have a different dynamical behaviour in a prominence. The differences have been detected at, both, large and small spatial scales. Large-scale ion-neutral drift velocities have been detected through the Doppler motion of the core of the observed spectral lines (\Halpha, \HeD3, and \CaII). The results show that ions have larger velocity excursions than neutrals by an amount usually smaller than 1 \kms. Small differences between the velocities of two the two neutral species have also been detected in particular moments indicating that the coupling between them is incomplete. On the other hand, small-scale motions have been probed through the width of the spectral lines. The analysis has led to the conclusion that ions also have more intense small-scale unresolved velocities than neutrals. With the present work, we cannot distinguish whether the additional required small-scale motions have a thermal or non-thermal nature. 

The more intense dynamical behaviour of ions has already been reported in previous works. For instance, \citet{Ramelli_etal_2012, Wiehr_etal_2013, Stellmacher_Wiehr_2015, Stellmacher_Wiehr_2017} obtained that the Doppler widths of spectral lines from ionised especies are, in general, larger than those from neutral atoms, although the relation did not always hold for all spectral lines of neutral species. For the large-scale velocities, \citet{Khomenko2016, Wiehr2019, Wiehr2021, Zapior2022} have detected higher values of the ions, but other works do not claim that behaviour \citep[for instance]{Anan2017}. 

A key aspect of our analysis may have been our selection criterion of the valid field-of-view. We have applied three constraints: (i) optically thin profiles (defined in our case by the amplitude of the \Halpha\ profile); (ii) asymmetric profiles were excluded (to make sure no velocity gradients along the line-of-sight existed); and (iii) multiple-peaked profiles were neither taken into account (to ensure the homogeneity of the measured plasma volume). Some of the previous works also applied some of these restrictions (especially the optically thin condition), but not all together, which might be the reason why we may have used a consistent sample of points to derive the trends in the analysed parameters, ensuring that every retained point represented a rather homogeneous plasma and that the observed spectral lines probed the same plasma volume. All those criteria have restricted the valid field-of-view to just the edge of the prominence. At these positions, close to the boundary between the prominence and the corona, a number of instabilities may take place as a consequence of the different physical conditions of the prominence and coronal plasma, helping to increase the differential ion-neutral behaviour and facilitate its detection. 

One point that deserves to be commented further is that ions tend to have larger velocities. In principle, the opposite might also be feasible, with neutrals trying to diffuse away from magnetic areas and ions keep trapped by the magnetic field. So, if ions move faster, we might conclude that this behaviour is a consequence of their response to the acceleration produced by magnetic forces and neutrals (insensitive to the Lorentz force directly) are not able to follow them. Unfortunately, we do not have information about the magnetic field and even less about its temporal fluctuations. So, a key conclusion of this work is that, if significant progress is to be made to understand the physical reason for the velocity drifts, the determination of the magnetic field in prominences is crucial (not only its average properties, but also its temporal variations so that we may be capable of understanding the spatial and temporal velocity changes). Some attempts have been done in the past to measure the magnetic field in prominences, but the results are still scarce \citep[see, v.g.,][]{Casini2003, Merenda2006, Orozco2014, Marian2015, Marian2016, Gibson2018}. The new generation of large solar telescopes \citep{DKIST2020, EST2022} will be most relevant to make those measurements feasible.

A second relevant conclusion refers to the acceptable field of view. Probably, the time has already come to start applying diagnostic tools that are not restricted to the optically thin case and that are more elaborated than the single or double slab scenario for the interpretation of non optically thin spectral lines, opening the door to include gradients in the physical parameters and facilitate the combination of spectral lines with different level of opacity and saturation effects. 
The non-LTE conditions that apply to prominences make the application difficult but a number of codes applied to on-disk data start to exist \citep[v.g.,][]{STIC2019, DeSIRe2022, TIC2022}.

Lastly, a third important conclusion refers to the spectral lines we have used in this work. Hydrogen and helium are light elements and their thermal velocities may be too large for a proper determination of the non-thermal motions. For a typical kinetic temperature of \hbox{10,000 K}, the thermal Doppler velocity of hydrogen and helium atoms is about 12 and 6 \kms, respectively. Small non-thermal Doppler velocities may be undetected with these spectral lines. On the other hand, calcium atoms are relatively heavy and their thermal motions (2 \kms\ for a kinetic temperature of \hbox{10,000 K}) may become irrelevant if non-thermal motions have large values as reported by some authors. The combination of hydrogen, helium, and calcium is appropriate because the effects of thermal and non-thermal velocities can be disentangled if they all share the same properties.  The inclusion of some additional spectral line of an atom with intermediate mass would help to separate the effect of thermal and non-thermal differential velocities between ions and neutrals. Unfortunately, the number of high-resolution telescopes that makes possible the simultaneous recording of several spectral lines is very limited. An effort to facilitate the possibility to have the required instrumentation set-ups would be welcome.


\begin{acknowledgements}
SJGM and EK are grateful for the support of the European Research
Council through the grant ERC-2017-CoG771310-PI2FA, and by the
Spanish Ministry of Science and Innovation through the grant PID2021-127487NB-I00. PG and SJGM acknowledge the support 
of the project VEGA 2/0048/20. Funding from the Horizon 2020 projects 
SOLARNET (No 824135) and ESCAPE (No 824064) is gratefully acknowledged. 
Financial support from the State Research Agency (AEI) of the Spanish 
Ministry of Science, Innovation and Universities (MCIU) 
and the European Regional Development Fund (FEDER) under grant with reference 
PGC2018-097611-A-I00 is gratefully acknowledged. This research was supported by the 
International Space Science Institute (ISSI) in Bern, through ISSI International Team
project 457 (The Role of Partial Ionization in the Formation, Dynamics
and Stability of Solar Prominences).
CK acknowledges funding from the European Union's Horizon 2020 research and innovation 
programme under the Marie Sk\l{}odowska-Curie grant agreement No 895955. TF acknowledges grant RYC2020-030307-I funded by MCIN/AEI/ 10.13039/501100011033 and by “ESF Investing in your future”
This research has made use of NASA's Astrophysics Data System.

\end{acknowledgements}


\bibliographystyle{aa}
\bibliography{aa-jour,biblio}









\end{document}